\documentclass[12pt]{article}
\usepackage{a4,epsf}

\def\thebibliography#1{\section*
 {References                                    % PLB/NPB style
 \markboth{REFERENCES}{REFERENCES}}\list
 {[\arabic{enumi}]}                             % PLB/NPB style
 {\settowidth\labelwidth{[#1]}\leftmargin\labelwidth
 \advance\leftmargin\labelsep
 \usecounter{enumi}}
 \def\newblock{\hskip .11em plus .33em minus -.07em}
 \sloppy \sfcode`\.=1000\relax}
 
\newcommand{\cleqn}{\setcounter{equation}{0}}
\newcommand{\clfn}{\setcounter{footnote}{0}}

\newcommand{\pr}{\hspace{\parindent}}

\newcommand{\bea}{\begin{eqnarray}}
\newcommand{\eea}{\end{eqnarray}}
\newcommand{\simgt}{\hbox{ \raise3pt\hbox to 0pt{$>$}\raise-3pt\hbox{$\sim$} }}
\newcommand{\simlt}{\hbox{ \raise3pt\hbox to 0pt{$<$}\raise-3pt\hbox{$\sim$} }}
\newcommand \vc[1]{{\bf {#1}}}

\def\to{\rightarrow}
\def\epem{\ifmmode{ e^{+}e^-} \else{$ e^{+}e^- $ } \fi}
\def\bw{\ifmmode{ bW^+ } \else{$ bW^+ $ } \fi}
\def\bwb{\ifmmode{ \bar{b}W^- } \else{$ \bar{b}W^- $ } \fi}
\def\ttbar{\ifmmode{t\bar{t}} \else{$t\bar{t}$} \fi}
\def\nrg{\ifmmode{\tilde{G}(p,E)} \else{$\tilde{G}(p,E)$} \fi}

\def\alpsmz{\alpha_{s}(m_Z)_{\overline{\bf MS}}}

\def \lsa {\rlap {\lower 3.5 pt \hbox {$\mathchar \sim$}} \raise 1
pt \hbox {$<$}}
\def \rsa {\rlap {\lower 3.5 pt \hbox {$\mathchar \sim$}} \raise 1
pt \hbox {$>$}}
\def\msbar {\ifmmode{\overline{\rm MS}}    \else{$\overline{\rm MS}$ }    \fi}
\def\oalfs {${\cal O}(\alpha_s) ~$}

\def\alpsmz{\ifmmode{\alpha_s(m_Z)_{\overline{\rm MS}}}
              \else{$\alpha_s(m_Z)_{\overline{\rm MS}}$} \fi}
\def\lamfive{\ifmmode{\Lambda^{(5)}_{\overline{\rm MS}}}
               \else{$\Lambda^{(5)}_{\overline{\rm MS}}$} \fi}
\def\lamfour{\ifmmode{\Lambda^{(4)}_{\overline{\rm MS}}}
               \else{$\Lambda^{(4)}_{\overline{\rm MS}}$} \fi}
\def\lsa{\rlap{\lower 3.5 pt \hbox {$\mathchar \sim$}} \raise 1pt \hbox {$<$}}
\def\rsa{\rlap{\lower 3.5 pt \hbox {$\mathchar \sim$}} \raise 1pt \hbox {$>$}}
\def\width{\Gamma_t}

\begin{document}
\begin{titlepage}
\title{Final-State Interactions in
$e^+e^- \to t\bar{t} \to bl^+\nu\bar{b}W^-$\\
Near Top Quark Threshold$^\dagger$
}
\author{ \\M. Peter$^\ddagger$ and Y. Sumino$^*$
\\
\\
{\normalsize \it Institut f\"ur Theoretische Teilchenphysik,
Universit\"at Karlsruhe,}\\
{\normalsize \it D-76128 Karlsruhe, Germany}}
\date{}
\maketitle
\thispagestyle{empty}
\vspace{-4.0truein}
\begin{flushright}
{\bf TTP 97-27}\\
{\bf hep-ph/9708223}\\
{\bf July 1997}
\end{flushright}
\vspace{3.0truein}
\vspace{3cm}
\begin{abstract}
\noindent
{\small
We calculate final-state interaction
corrections to the energy-angular distribution of $l^+$ in semi-leptonic
top quark decay, where the parent top quark is
produced via $e^+e^- \to t\bar{t}$ near threshold.
These are the corrections due to gluon exchange between
$t$ and $\bar{b}$ ($\bar{t}$ and $b$) and between
$b$ and $\bar{b}$.
Combining with previously known other corrections, we explicitly
write down the $l^+$ energy-angular distribution 
including the full ${\cal O}(\alpha_s)={\cal O}(\beta )$ corrections
near $t\bar{t}$ threshold.
Numerical analyses of the final-state interaction corrections
are given.
We find that they deform the $l^+$ distribution 
typically at the 10\% level.
We also find that all qualitative features of the numerical results can be
understood from intuitive pictures.
The mechanisms of various effects of the final-state
interactions are elucidated.
Finally we define an observable which 
is proper to the decay process of the top quark
(dependent only on $d\Gamma_{t \to bl^+\nu}/ dE_l d\Omega_l$
of a free polarized top quark) near $t\bar{t}$ threshold.
Such a quantity will be useful in extracting the decay properties of 
the top quark using the highly polarized top quark samples.
}
\end{abstract}
\vfil

\vspace{2cm}
\footnoterule
\footnotesize
\vspace{3mm}
\hspace{-6mm}
$^\dagger$
Work supported in part by Graduiertenkolleg ``Elementarteilchenphysik
an Beschleunigern'', by the ``Landesgraduiertenf\"orderung'' at the
University of Karlsruhe, by BMBF under contract 057KA92P, and by
the Alexander von Humboldt Foundation.

\hspace{-6mm}
$^\ddagger$
Address after October 1, 1997: Institut f\"ur Theoretische Physik,
Universit\"at Heidelberg, Philosophenweg 16, D-69120 Heidelberg, Germany.

\hspace{-6mm}
$^*$ On 
leave of absence from Department of Physics, Tohoku University,
Sendai 980-77, Japan.
\end{titlepage}
%

%%%%% text
%\baselineskip 22pt

\section{Introduction}
\cleqn
\clfn

A future $e^+e^-$ linear collider operating at energies
around the $t\bar{t}$ threshold will be one of the
ideal testing grounds for
unraveling the properties of the top quark. So far there have been a number
of studies of the cross section for top-quark pair production
near the $t\bar{t}$ threshold, both
theoretical and experimental \cite{r7}--\cite{r26}, in which it has been
recognized
that this kinematical region is rich in physics and is also apt for extracting
various physical parameters efficiently, e.g.\ $m_t$, $\alpha_s$, $\Gamma_t$,
$m_H$, $g_{tH}$, etc.

While most of the previous analyses were solely concerned with the production
process of the top quark, one may also analyze the decay process
in detail and extract some important physics information.
Especially the fact that
$t$ and $\bar{t}$ are produced highly polarized in the
threshold region is potentially quite advantageous 
for studying the electroweak properties of the top quark through its decay.
Detailed investigations of the decay of free polarized top quarks have
already been available including the full ${\cal O}(\alpha_s)$ 
corrections \cite{r27,r28}.
Close to threshold, however, these precise
analyses do not apply directly
because of the existence of corrections unique to this region,
which connect the production and decay
processes of the top quark.
Specifically, these are 
the corrections due to gluon exchange between $t$
and $\bar{b}$ ($\bar{t}$ and $b$) or between $b$ and $\bar{b}$.

This type of corrections arises when the particles produced decay quickly 
into many particles, and are referred to
as ``final-state interactions'', ``rescattering corrections'',
or ``non-factorizable corrections'' in the literature.
They generally vanish in inclusive cross 
sections \cite{r21,r40,r22},
but modify the shape of differential distributions 
in a non-trivial manner \cite{r40,cracow,r26}.
The size of the corrections is at the 10\% level 
in the $t\bar{t}$ threshold region, hence it is 
inevitable to incorporate their
effects in precision studies of top-quark production and decay near
threshold.
The same kind of effects has recently been studied in $W$ 
pair-production \cite{my,khst,khsj,bcb}.

The first analysis of the top-quark decay in the threshold
region was given as a part of the results in Ref.\ \cite{r26}. In that paper,
the mean value $\langle n \ell \rangle$
of the charged lepton four-momentum projection on an arbitrarily chosen
four-vector $n$ in semi-leptonic top decays was proposed as an experimentally
observable quantity sensitive
to top quark polarization, and this quantity was calculated
including the final-state interactions.
Clearly, and also admittedly in that work, the calculation of
the differential distribution of $l^+$ including the
final-state interactions has been demanded.

In this paper, we calculate these final-state interaction corrections to the
differential energy-angular distribution of the charged leptons.
% arising from
%the semi-leptonic top quark decay in the $t\bar{t}$ threshold region.
We find that the corrections deform the $l^+$ distribution non-trivially at
the expected level. Combining with other, previously known results,
we write down
the explicit formula for the $l^+$ energy-angular distribution including
all ${\cal O}(\alpha_s)={\cal O}(\beta )$ corrections near
$t\bar{t}$ threshold.

Another aim of this paper is to present
physical descriptions of the final-state interactions which enable us 
to qualitatively understand the features of our numerical results.
Such descriptions
would be useful since
the systematic calculation of final-state interactions
based on quantum field theory is
rather complicated, involving box- and pentagon-type diagrams, and 
it is not easy to make physical sense
out of the obtained final expressions.
%It turns out that one may explain various effects of the
%final-state interactions on the
%differential distributions by following the classical picture
%that there is
%an attractive force between $t$ and $\bar{b}$ (or between $\bar{t}$ and $b$).
To our knowledge such qualitative explanations have never 
been put forth, although corresponding theoretical calculations 
and numerical studies have been partly available.

Finally we propose an observable which is proper to the decay process of
the top quark near threshold.
Since the final-state interaction
connects the production and decay processes
of the top quark, it destroys the factorization property of the corresponding
production and decay cross sections.
In order to study the decay of the top quark in a clean environment, 
it would be useful if we could find an observable which depends
only on this process
($d\Gamma_{t \to bl^+\nu}/ dE_l d\Omega_l$
of a free polarized top quark).
In fact, such an observable can be constructed, which
at the same time preserves most of the 
differential information of the $l^+$ energy-angular distribution.

In our numerical analysis
we solve the Schr\"{o}dinger equation numerically in order to include the
QCD binding effects near threshold.
We follow both the coordinate-space approach developed in Refs.\ \cite{r8,r9}
and the momentum-space approach developed in Refs.\ \cite{r35,r36}
in solving the equation, and compare the results.
There are small differences in the numerical results obtained
from the two approaches,
reflecting the difference in the construction of the potentials at short
distance. 
(The difference is formally ${\cal O}(\alpha_s^2)$, of the order beyond 
our present scope.) This issue is also discussed.

In section 2 we introduce some notations to be used in later 
sections.
Section 3 contains the physical descriptions of the effects of
final-state interactions.
The results of the systematic calculation of final-state interaction
corrections to the $l^+$ energy-angular distribution, as well as
the complete formula for the distribution including all 
${\cal O}(\alpha_s)={\cal O}(\beta )$ corrections, are given in section 4.
Section 5 shows various numerical results and a comparison with the
qualitative picture.
We define an observable proper to the top decay process in section 6.
Discussion and conclusion are presented in sections 7 and 8, respectively.

\section{Definitions and Conventions}
\cleqn
\clfn

We consider longitudinally polarized \epem beams throughout
our analyses.
$P_{e^{\pm}}$ denotes the longitudinal polarization of $e^{\pm}$,
and we set
\begin{equation}\label{chi}
\chi={P_{e^+}-P_{e^-}\over1-P_{e^+}P_{e^-}} .
\end{equation}
We choose a reference coordinate system in the \ttbar c.m.\ frame.
Three orthonormal basis vectors are defined as
%\begin{eqnarray}
\begin{equation}
\hat{\bf n}_{\|} = \frac{{\bf p}_{e^-}}{|{\bf p}_{e^-}|}\quad,\quad%\nonumber\\
\hat{\bf n}_{\rm N} = {{\bf p}_{e^-} \! \times {\bf p}_t, \over
        |{\bf p}_{e^-} \! \times {\bf p}_t|} \quad,\quad%              \label{basis}\\
\hat{\bf n}_\bot = \hat{\bf n}_{\rm N} \times \hat{\bf n}_{\|} ~,  
 \label{basis} %\nonumber
%\end{eqnarray}
\end{equation}
where $\vc{p}_{e^-}$ and $\vc{p}_t$ represent the $e^-$ and $t$ momentum,
respectively. 

Our conventions for the fermion
vector and axial-vector couplings to the $Z$ boson are
\begin{equation}
v_f = 2 I^3_f - 4 q_f \sin^2\theta_{\rm W} , \qquad  a_f = 2 I^3_f ,
\end{equation}
respectively.
Certain combinations of these couplings will be useful below:
\begin{eqnarray}
a_1 &=& q_e^2 q_t^2 + (v_e^2 + a_e^2) v_t^2 d^2 +
        2 q_e q_t v_e v_t d ,\nonumber \\
a_2 &=& 2 v_e a_e v_t^2 d^2 + 2 q_e q_t a_e v_t d ,\nonumber \\
a_3 &=& 4 v_e a_e v_t a_t d^2 + 2 q_e q_t a_e a_t d ,\label{coupl}\\
a_4 &=& 2 (v_e^2 + a_e^2) v_t a_t d^2 + 2 q_e q_t v_e a_t d, \nonumber\\
d &=& {1\over 16 \sin^2\theta_{\rm W}\cos^2\theta_{\rm W}}\,{s\over s - M_Z^2},
    \nonumber
\end{eqnarray}
and 
\\
\parbox{75.ex}{
\begin{eqnarray*}
& &\hspace{5.ex}C_\|^0 (\chi) =
-{a_2 + \chi a_1 \over a_1 + \chi a_2} ,\hspace{6.9ex}
  C_\|^1 (\chi) = \left( 1-\chi^2 \right) {a_2 a_3 - a_1 a_4   \over
        \left(a_1 + \chi a_2 \right)^2} ,\\
& &\hspace{5.ex}C_\bot(\chi)  = -{1\over 2} \,
{a_4 + \chi a_3 \over a_1 + \chi a_2} ,
    \qquad C_{\rm N}(\chi) =-{1 \over 2}\, {a_3
    + \chi a_4 \over a_1 + \chi a_2}\, =\, - C_{\rm FB}(\chi) .
\end{eqnarray*}}
\hfill
\parbox{5.ex}{
\begin{eqnarray} \label{coefs} \end{eqnarray} }\\
Finally, we define 
\bea
y = \frac{M_W^2}{m_t^2}, ~~~~~~~~
\kappa = \frac{1-2y}{1+2y}.
\label{ykappa}
\eea

The QCD enhancement of the top-quark production cross section 
near threshold is incorporated through the
$S$-wave and $P$-wave Green's functions, $\nrg$ and $\tilde{F}(p,E)$,
of the non-relativistic Schr\"{o}dinger equations in the presence
of the QCD potential.
These functions are defined by
\bea
&&
\left[
- \frac{\nabla^2}{m_t} + V_{QCD}(r) -
\left( E+i\Gamma_t \right)
\right] \, G(\vc{x},E) = \delta^3(\vc{x}),
\label{sse2}
\\
&&
\left[
- \frac{\nabla^2}{m_t} + V_{QCD}(r) -
\left( E+i\Gamma_t \right)
\right] \, F^k(\vc{x},E) = -i \partial^k \delta^3(\vc{x}),
\label{pwse2}
\eea
and
\bea
\tilde{G}(p,E) &=& \int d^3\vc{x} \, e^{-i\vc{p} \cdot \vc{x} } \, 
G(\vc{x},E) ,
\\
p^k \,
\tilde{F}(p,E) &=& \int d^3\vc{x} \, e^{-i\vc{p} \cdot \vc{x} } \, 
F^k(\vc{x},E) .
\label{ftpgf2}
\eea
One may obtain the Green's functions either by first solving the 
Schr\"{o}dinger 
equations in coordinate space and taking the Fourier transforms of the
solutions \cite{r9},
or by solving the Schr\"{o}dinger equations directly in
momentum space \cite{r35}.

Various known \oalfs corrections to threshold cross sections can also be 
expressed  in terms of the above Green's functions.
In Ref.\ \cite{r26}
the following functions have been defined, which are
solely determined from QCD, to represent various independent
corrections.
We will pursue the same conventions in this paper.
\bea
&&
\varphi(p,E) =
{(1-{C_F \alpha_{s}(m_t)/ \pi})\over (1-{2 C_F \alpha_{s}(m_t)/ \pi})}\,
        {p \over m_t}\,
        {\tilde{F}^* \!(p,E) \over \tilde{G}^* \!(p,E)}  ,
\\
&&
\varphi_{\rm _R} = \mbox{Re}\,\varphi , ~~~~~
\varphi_{\rm _I} = \mbox{Im}\,\varphi ,
\\
&&
\psi_1(p,E) = - \, C_F \! \cdot \! 4\pi \alpha_s (\mu_B )
\int \! \mbox{$\frac{d^3\vc{q}}{(2\pi)^3}$} \,
\frac{1 \,}{|\vc{q}\! -\! \vc{p}_t|^3} \,
2 \, \mbox{Im}
\biggl[ \frac{\tilde{G}(q,E)}{\tilde{G}(p,E)} \biggl] 
\cdot \frac{\pi}{2}
\label{psi1}
\\
&&
\psi_{\rm _R}(p,E) = - \, C_F \! \cdot \! 4 \pi \alpha_s (\mu_B ) \,
\mbox{Pr.}\int \mbox{$\frac{d^3\vc{q}}{(2\pi)^3}$} \,
\, \frac{1 \,}{|\vc{q}\! -\! \vc{p}_t|^3} \,
\frac{\vc{p}_t\! \cdot \! (\vc{q}\! -\! \vc{p}_t)}{|\vc{p}_t|\, 
|\vc{q}\! -\! \vc{p}_t|}
\,
2 \,
\mbox{Re}
\biggl[ \frac{\tilde{G}(q,E)}{\tilde{G}(p,E)} \biggl] .
\label{psi2}
\eea
Here $C_F=4/3$ is a color factor; $p=|\vc{p}_t|$ and $q=|\vc{q}|$;
$\mu_B$ is taken to be 15~GeV in our analyses.
Eqs.\ (\ref{psi1}) and (\ref{psi2})
differ slightly in their integrands from those defined
in Ref.\ \cite{r26}.\footnote{
In Ref.\ \cite{r26} the Coulomb propagator between $b$ and $\bar{t}$
(or between $\bar{b}$ and $t$) together with the color charges
are replaced by the QCD potential
between two heavy quarks by hand,
$-C_F \! \cdot \! 4\pi\alpha_s/|\vc{q}-\vc{p}_t|^2 \to 
\tilde{V}_{QCD}(|\vc{q}-\vc{p}_t|)$.
One advantage of Eqs.\ (\ref{psi1}) and (\ref{psi2}) is that
one may convert them into one-parameter integral forms by explicitly 
integrating over $d\Omega_{\vc{q}}$ \cite{r40,r18}.
}
The differences, however, can be regarded as higher order corrections
which are beyond the scope of our analysis.

\section{Qualitative Description of Final-State Interaction Effects}
\cleqn
\clfn

In this section a qualitative understanding of the 
effects of final-state interactions is explained.
It is based on the classical picture that 
$t$ and $\bar{b}$ ($\bar{t}$ and $b$) attract
each other due to their colour charges.
We will see that all the following qualitative features 
match well with our numerical
studies presented in section 5.
Moreover, the following argument 
will help interpreting the formulas
derived in the next section.

\subsection{Top Momentum Distribution}
\begin{figure}
\begin{center}
  \leavevmode
  \epsfxsize=5.cm
  \epsffile{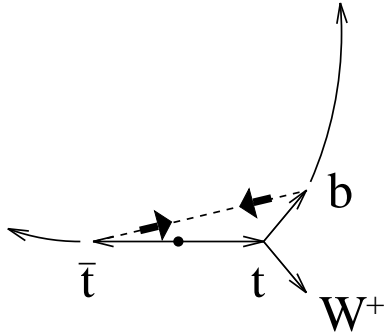}\\
  \caption[]{
        \label{forcebtbar}Attractive Coulomb force between $\bar{t}$
        and $b$ (from $t$ decay). The momentum transfer 
        $\delta \vc{p}_b = - \delta \vc{p}_{\bar{t}}$ due to
        the attraction is indicated by thick arrows.}
\end{center}
\end{figure}

Let us first consider the effect of final-state interactions on
the top-quark momentum ($|\vc{p}_t|$) distribution, where
$\vc{p}_t$ is reconstructed from the $bW^+$ momenta at time
$\tau \to \infty$.
The energy of a $t\bar{t}$ system before its decay is given by
\bea
{\cal E} = E_t + E_{\bar{t}} + V( |\vc{r}_t - \vc{r}_{\bar{t}}|),
\eea
where
\bea
E_t = m_t + \frac{\vc{p}_t^2}{2m_t}, \qquad
E_{\bar{t}} = m_t + \frac{\vc{p}^2_{\bar{t}}}{2m_t}, \qquad
V(r)=-C_F \frac{\alpha_s}{r}.
\eea
Suppose $t$ decays first and 
let $a \sim (\alpha_s m_t)^{-1} \sim (m_t \Gamma_t)^{-1/2}$ be 
the typical distance between $t$ and $\bar{t}$ at the time of $t$ decay.
Then, just before the decay, the momenta of $t$ and $\bar{t}$ are given by
\bea
|\vc{p}_t| = |\vc{p}_{\bar{t}}| \sim
\sqrt{
m_t \left[ {\cal E} - 2m_t + |V(a)| \, \right]
} .
\eea
Their order of magnitude is 
$\alpha_s m_t \sim (\mbox{Bohr radius})^{-1}$.
If it were not for the final-state interactions between $\bar{t}$ and $b$,
and if $\bar{t}$, $b$ and $W^+$ traveled as free particles, 
the above momentum would be transferred to
the $bW^+$ system at time
$\tau \to \infty$:
\bea
|\vc{p}_b + \vc{p}_{W^+}|_{\tau \to \infty} \sim
\sqrt{
m_t \left[ {\cal E} - 2m_t + |V(a)| \, \right]
} .
\eea
Taking into account the final-state interaction
(Coulomb interaction) between $\bar{t}$ and $b$, 
the energy of the $\bar{t}bW^+$ system is given by\footnote{
Here we neglect the interaction of $\bar{t}$ ($b$) and 
the magnetic field generated by $b$ ($\bar{t}$).
This approximation is justified in the case of our interest;
see section 4.
}
\bea
{\cal E} = E_b + E_{W^+} + E_{\bar{t}} +
V( |\vc{r}_b - \vc{r}_{\bar{t}}|) ,
\\
E_b = |\vc{p}_b|, ~~~~~ E_{W^+} = \sqrt{ \vc{p}_{W^+}^2 + M_W^2 } .
\eea
As depicted in Fig.~\ref{forcebtbar}, the tracks of $b$ and $\bar{t}$
are deflected due to the attraction between the two particles, which
lose kinetic energy as $b$ flies off to infinity at the
speed of light.
The classical equation of motion is given by
\bea
\frac{d\vc{p}_b}{d\tau} = - \frac{d\vc{p}_{\bar{t}}}{d\tau}
= - \frac{\partial}{\partial \vc{r}_b} \, 
V( |\vc{r}_b - \vc{r}_{\bar{t}}|) .
\eea
Substituting the free particle solution 
$\vc{r}_b = \vc{v} \tau + \vc{r}_0$ on the right-hand side and
noting $|\vc{r}_b - \vc{r}_{\bar{t}}| \simeq \tau$,
we may estimate the size of the momentum transfer due to the attractive force 
as\footnote{
It corresponds to solving the equation of motion by a series expansion
of $\alpha_s$.
}
\bea
|\delta \vc{p}_b| = |\delta \vc{p}_{\bar{t}}| \sim
|V( r_{min})| ,
\label{momtr}
\eea
where the minimum distance between $b$ and $\bar{t}$ is denoted as
$r_{min}$.
Typically $r_{min} \sim a$, and
we find from Eq.\ (\ref{momtr}) that the effect of final-state interactions
on the $|\vc{p}_t|$ distribution is of the order of
$|\delta \vc{p}_t| \sim \alpha_s^2 m_t$.
Obviously the effect is to reduce $|\vc{p}_t|$.

\subsection{Forward-Backward Asymmetric Distribution}
Next we consider the $\cos \theta_{te}$ distribution of the top quark.
($\theta_{te}$ denotes the angle between $t$ and $e^-$ 
in the \ttbar c.m.\ frame.)
It has been known that a forward-backward asymmetric distribution
of the top quark is generated by the final-state 
interactions \cite{r40,r26}.
We describe its mechanism here.

\begin{figure}
\begin{center}
  \leavevmode
  \epsfxsize=7.cm
  \epsffile{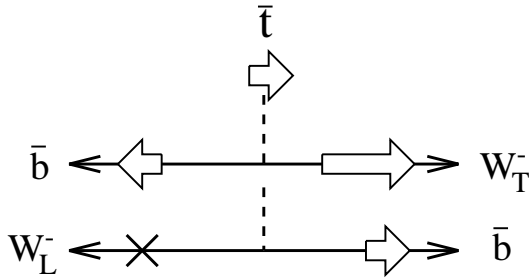}\\
  \caption[]{
        \label{tbardecay}Typical configurations in the decay 
        of $\bar{t}$ with definite
        spin orientation.  Transverse $W^-$ ($W^-_T$) tend 
        to be emitted in the direction of the $\bar{t}$ spin orientation,
        while longitudinal $W^-$ ($W^-_L$) are emitted 
        in the opposite direction due to helicity conservation.
        For $m_t \simeq 175$~GeV, $\bar{t}$ decays mainly to
        $W^-_L$, hence $\bar{b}$ is emitted more in the $\bar{t}$ spin
        direction.}
\end{center}
\end{figure}

\begin{figure}
\begin{center}
  \leavevmode
  \epsfxsize=12.cm
  \epsffile{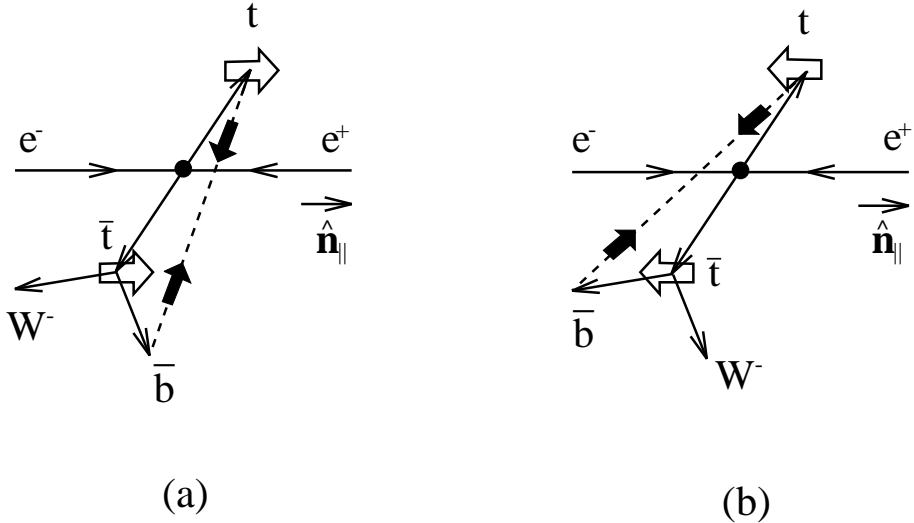}\\
  \caption[]{
        \label{spin}Attractive force between $t$ and $\bar{b}$
        when the $t$ and $\bar{t}$ spins are oriented in the
        (a) $\hat{\vc{n}}_\|$ direction, and in the
        (b) $-\hat{\vc{n}}_\|$ direction. The 
        momentum transfer  
        $\delta \vc{p}_b = - \delta \vc{p}_{\bar{t}}$ due to
        the attraction is indicated by thick black arrows.}
\end{center}
\end{figure}

We consider the case where $\bar{t}$ decays first and examine the
interaction between $t$ and $\bar{b}$.
The $t$ and $\bar{t}$ pair-produced near threshold in $e^+e^-$ collisions
have their spins approximately parallel or anti-parallel to the $e^-$ beam
direction and
the spins are always oriented parallel to each other.
In fact in leading order the polarization vector of the top quark is
given by \cite{kuehn,r38}
$\mbox{\boldmath$\cal P$} = C_\|^0(\chi) \, \hat{\vc{n}}_\|$.
On the other hand, the decay of $\bar{t}$ occurs via a $V \! -\! A$
coupling, and
$\bar{b}$ is emitted preferably in the spin direction of the parent $\bar{t}$,
see Fig.\ \ref{tbardecay}.
More precisely,
the excess of the $\bar{b}$'s 
emitted in the $\bar{t}$ spin direction over those
emitted in the opposite direction is given by 
$\kappa$ defined in Eq.\ (\ref{ykappa}).
Now suppose $t$ and $\bar{t}$ have their spins in the $\hat{\vc{n}}_\|$
direction.
Then $\bar{b}$ will be emitted dominantly in the $\hat{\vc{n}}_\|$
direction.
One can see from Fig.~\ref{spin}(a)
that in this case $t$ is always attracted to the forward direction 
due to the attractive force between $t$ and $\bar{b}$.
The direction of the attractive force will be opposite
if $t$ and $\bar{t}$ have
their spins in the $-\hat{\vc{n}}_\|$ direction (Fig.~\ref{spin}(b)).
Thus, polarized top quarks will be pulled in a
definite (forward or backward) direction,
and we may expect 
that a forward-backward asymmetric distribution of
the top quark
$\sim \kappa \, C_\|^0(\chi) \cos \theta_{te}$ is
generated by the final-state interaction.

Incidentally, a forward-backward asymmetric distribution of the top quarks
is also generated by the interference
between the $S$-wave and $P$-wave $t\bar{t}$ pair-production 
amplitudes \cite{r11}.
It is formally a quantity of the same order as the final-state
interaction near threshold, since it arises as an 
${\cal O}(\beta )={\cal O}(\alpha_s)$ correction to the leading
spherically symmetric distribution.
Interestingly, 
we find that each $\cos \theta_{te}$ distribution
has quite a different physical explanation for
its generation mechanism, although it is a common feature that both 
originate from the interplay between QCD and electroweak interactions.

\subsection{Top Quark Polarization Vector}
The Coulomb attraction between $t$ and $\bar{b}$ also modifies
the top-quark polarization vector \cite{r26}.
As we have seen previously, in the $\bar{t}$ decay,
$\bar{b}$ tends to be emitted in the direction of the parent
$\bar{t}$'s spin direction (Fig.\ \ref{tbardecay}).
We then find from Fig.\ \ref{spin} 
that if the $t$ and $\bar{t}$ spins are oriented in the $\hat{\vc{n}}_\|$ 
direction, $t$ will be attracted to the forward direction
due to the attraction by $\bar{b}$, and oppositely attracted
to the backward direction if the $t$ and $\bar{t}$ spins are 
in the $-\hat{\vc{n}}_\|$ direction.
This means that in the forward region ($\cos \theta_{te} \simeq 1$)
the number of $t$'s with
spin in $\hat{\vc{n}}_\|$ direction increases whereas in the
backward region the number of 
those with spin in the opposite direction increases.
Or equivalently, the $\hat{\vc{n}}_\|$-component of
the top-quark polarization vector
increases in the forward region and decreases in the backward region.
We may thus conjecture that the top-quark polarization vector is modified
as 
$\delta \mbox{\boldmath$\cal P$} \sim \kappa \cos \theta_{te} \, 
\hat{\vc{n}}_\|$
due to the interaction between $t$ and $\bar{b}$.

\subsection{\boldmath $l^+$ Energy-Angular Distribution}
\begin{figure}
\begin{center}
  \leavevmode
  \epsfxsize=12.cm
  \epsffile{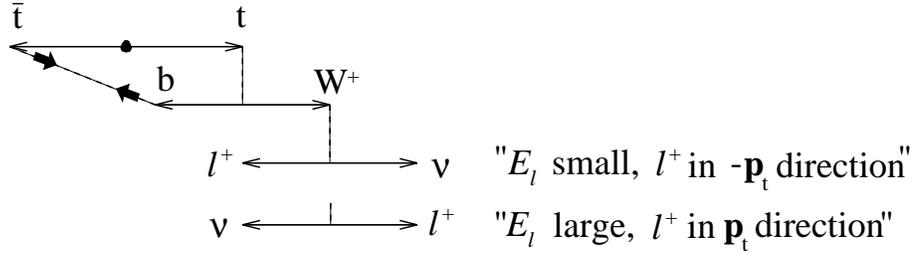}\\
  \caption[]{
        \label{semildecay}Typical configurations of the particles
        in semi-leptonic decay of $t$ when the $b$-quark
        is emitted in the $\bar{t}$ direction.
        Due to the boost by $W^+$, the energy-angle correlation of $l^+$
        will be either
``$E_l$ is small and $l^+$ emitted in $- \vc{p}_t$ direction'' or
``$E_l$ is large and $l^+$ emitted in $\vc{p}_t$ direction''.}
\end{center}
\end{figure}

Finally let us examine the effect of the Coulomb attraction between
$b$ and $\bar{t}$ on the $l^+$ energy-angular distribution in the
semi-leptonic decay of $t$.
The $b$-quark from $t$ decay
will be attracted in the direction of $\bar{t}$ due to
the Coulomb interaction between these two particles.
We show schematically typical configurations of the particles
in the top-quark semi-leptonic decay in Fig.\ \ref{semildecay}.
It can be seen that if the probability for $b$ being emitted in
the $\bar{t}$ direction increases, correspondingly
the probability for particular $l^+$ energy-angular configurations 
increases.
These configurations are either
``$E_l$ is small and $l^+$ emitted in $- \vc{p}_t$ direction'' or
``$E_l$ is large and $l^+$ emitted in $\vc{p}_t$ direction''.

\section{\boldmath $l^+$ Energy-Angular Distribution}
\cleqn
\clfn

We present the formulas for the charged lepton energy-angular distribution
in the decay of a top quark that is produced via $\epem \to \ttbar$
near threshold.
In the following, $x_l=2E_l/m_t$ and $\Omega_l$, respectively, denote
the normalized energy and the solid angle of the charged lepton as defined in
the rest frame of the parent top quark.
For simplicity we neglect the decay of $W^-$ in our calculations.

\subsection{Factorizable Part}

\begin{figure}
\begin{center}
  \leavevmode
  \epsfxsize=8.cm
  \epsffile{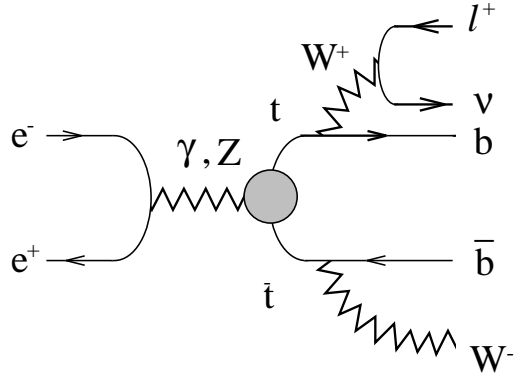}\\
  \caption[]{
        \label{bornfig}Born-type diagram for the process
$e^+e^- \to t\bar{t} \to bl^+\nu\bar{b}W^-$.
}
\end{center}
\end{figure}
It is well-known that the
contribution of the Born-type (reducible) diagram (Fig.~\ref{bornfig}) 
to the
differential distribution of $t$ and $l^+$ has a form where the
production and decay processes of the top quark
are factorized:
\bea
\frac{d\sigma_{\mbox{\scriptsize{Born}}}(\epem \! \to \ttbar \to bl^+\nu\bar{b}W^-)}
{d^3\vc{p}_t dx_l d\Omega_l}
=
\frac{d\sigma_{\mbox{\scriptsize{Born}}}(e^+e^- \! \to t\bar{t})}
{d^3\vc{p}_t} \times
\frac{1}{\Gamma_t} \,
\frac{d\Gamma_{t \to bl^+\nu}
(\mbox{\boldmath$\cal P$}_{\mbox{\scriptsize{Born}}}) }
{dx_l d\Omega_l} .
\label{born}
\eea
The above form holds true even including 
${\cal O}(\alpha_s)={\cal O}(\beta )$ corrections to each vertex
and propagator in the Born-type diagram.
Here, $\width$ \cite{r42,r53} and
$d\Gamma_{t \to bl^+\nu}(\mbox{\boldmath$\cal P$}_{\mbox{\scriptsize{Born}}})
/ dx_l d\Omega_l$ \cite{r44,r27},
respectively, are the width of a free top quark and
the charged lepton ($l^+$) energy-angular distribution in
the decay of a free polarized top quark,
both including the corresponding full \oalfs corrections.
Near threshold, the top-quark production cross section for longitudinally
polarized \epem beams is given by \cite{r38,r26}
\bea
&&
\frac{d\sigma_{\mbox{\scriptsize{Born}}}(e^+e^- \! \to t\bar{t})}
{d^3{\bf p}_t} 
= \frac{d\sigma^0_{t\bar{t}}}
{d^3{\bf p}_t} \times
\left[
1 + 2 C_{FB}(\chi ) \varphi_{\rm _R}(p,E) \cos \theta_{te}
\right] ,
\\
&&
\frac{d\sigma^0_{t\bar{t}}}
{d^3\vc{p}_t} 
= \frac{N_c\alpha^2\Gamma_t}{4\pi m_t^4} \,
\left( 1 - P_{e^+}P_{e^-} \right) ( a_1 + \chi a_2 )
\left( 1 - \frac{4C_F\alpha_s(m_t)}{\pi} \right)
\left| \nrg \right|^2 .
\label{dstt0}
\eea
Here, $\alpha$ is the fine structure constant, and $N_c=3$.
$\mbox{\boldmath$\cal P$}_{\mbox{\scriptsize{Born}}}
= {\cal P}_\| \hat{\vc{n}}_\| + {\cal P}_\bot \hat{\vc{n}}_\bot + 
{\cal P}_{\rm N} \hat{\vc{n}}_{\rm N}$ 
represents the polarization vector of a top quark produced via the
Born-type diagram (Fig.~\ref{bornfig}) near threshold.
The components are given by
\begin{eqnarray}
{\cal P}_\|(\vc{p}_t ,E,\chi) &=& C_\|^0(\chi)
+ C_\|^1(\chi)\, \varphi_{\rm _R}(p ,E)\,\cos\theta_{te}\,
 \label{thr_long}\\
{\cal P}_\bot(\vc{p}_t ,E,\chi) &=& C_\bot(\chi)\,
\varphi_{\rm _R}(p ,E)\,
\sin\theta_{te}\,
\label{thr_perp}\\
{\cal P}_{\rm N}(\vc{p}_t ,E,\chi) &=& C_{\rm N}(\chi)
\varphi_{\rm _I}(p ,E)
\sin\theta_{te}\,
        \label{thr_norm} .
\end{eqnarray}

Let us review briefly how
to derive the factorized form of the differential
distribution, Eq.(\ref{born}). 
First, in calculating the fully-differential cross section 
$d\sigma /d\Phi_5(bl^+\nu\bar{b}W^-)$, 
one may replace the top-quark momentum by an on-shell four-vector as
\bea
p_t^\mu = (p_t^0,\vc{p}_t) ~~~ \to ~~~
\tilde{p}_t^\mu = (\sqrt{\vc{p}_t^2+m_t^2},\vc{p}_t)
\label{replpt}
\eea
in vertices and propagator numerators
(but not in propagator denominators).
The replacement is justified because near threshold relevant
kinematical configurations are determined by
\bea
p_t^0 -m_t \sim \alpha_s^2 m_t, ~~~~~
|\vc{p}_t| \sim \alpha_s m_t
\eea
so that the replacement induces differences only at ${\cal O}(\alpha_s^2)$,
and also because we will not be concerned with the $p_t^0$-dependence of the
cross section.
Then, one may use the following identity
to factorize the spinor traces that appear in the fully-differential
cross section into
their production and decay parts:
{\it For an arbitrary $4 \times 4$ spinor matrix $G$ and for a
four-vector $\tilde{p}_t^\mu$ satisfying $\tilde{p}_t^2 = m_t^2$,}
\bea
\frac{\not \! \tilde{p}_t + m_t}{2m_t} \, G \,
\frac{\not \! \tilde{p}_t + m_t}{2m_t}
= \frac{\not \! \tilde{p}_t + m_t}{2m_t} \, 
\frac{1-\not \! {\cal P}\gamma_5}{2} \,
C ,
\label{id1}
\eea
{\it where the four-vector ${\cal P}^\mu$ and the constant $C$ are determined
from $G$ and $\tilde{p}_t^\mu$ via the relation}
\bea
\frac{1-{\cal P}\! \cdot\! s}{2} \, C =
\mbox{Tr} \left[
\frac{\not \! \tilde{p}_t + m_t}{2m_t} \, G \,
\frac{\not \! \tilde{p}_t + m_t}{2m_t} \,
\frac{1-\not \! s\gamma_5}{2} \right] ,
\label{id2}
\eea
{\it provided $s^\mu$ and ${\cal P}^\mu$ satisfy 
$s \cdot \tilde{p}_t = {\cal P}\cdot \tilde{p}_t =0$.}

One may also factorize the phase-space as
\bea
&&
d\Phi_5(\gamma^* \to bl^+\nu\bar{b}W^-) = 
\frac{d^4p_t}{(2\pi)^4} \, d\Phi_3(t^* \to bl^+\nu) \, 
d\Phi_2(\bar{t}^* \to \bar{b}W^-) ,
\\
&&
d\Phi_3(t^* \to bl^+\nu) = \biggl( \frac{1}{4\pi} \biggl)^{\! 5}
dx_l d\Omega_l dp_W^2 d\phi_{bl},
\label{decphsp}
\eea
where $p_W^2$ is the invariant-mass-squared of $l^+\nu$, and
$\phi_{bl}$ denotes the azimuthal angle
of $b$ around $l^+$ in the top-quark rest frame.
Then one integrates over $dp_W^2$, $d\phi_{bl}$, 
$d\Phi_2(\bar{b}W^-)$ and $dp_t^0/(2\pi )$;
the integration over $p_W^2$ is trivial since
we use the narrow-width approximation for $W^+$; the 
integration over $d\phi_{bl}$ is also trivial since the
fully-differential cross section
is independent of $\phi_{bl}$; the integration over the
$\bar{b}W^-$ phase-space
merely replaces the $\bar{b}W^-$ wave functions by $\width$; 
the $p_t^0/(2\pi )$-integration is straightforward.

\subsection{Final-State Interaction Corrections}
\begin{figure}
\begin{center}
  \leavevmode
  \epsfxsize=16.cm
  \epsffile{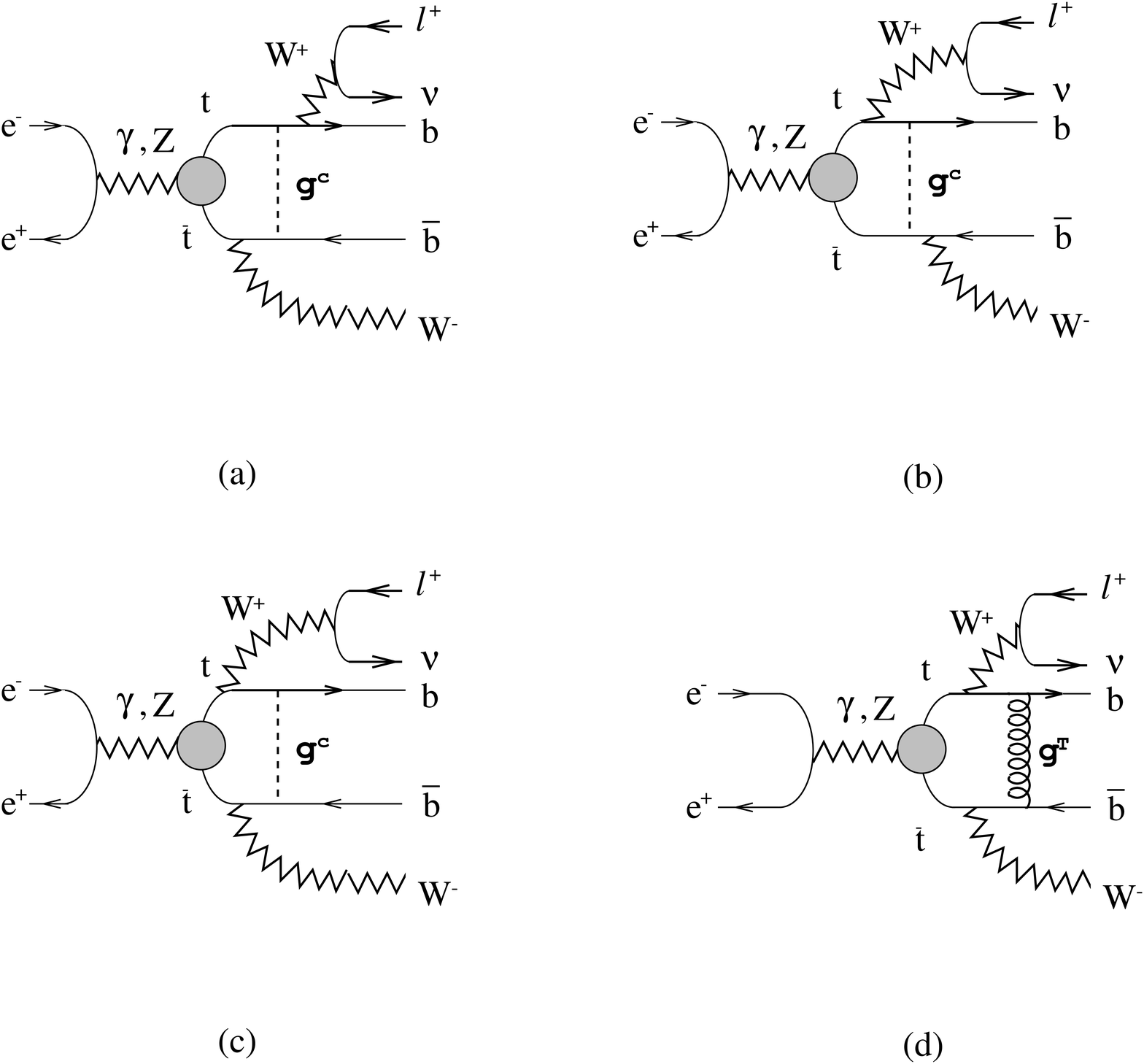}\\
  \caption[]{
        \label{fsifig}
Diagrams for final-state interactions for 
$e^+e^- \to t\bar{t} \to bl^+\nu\bar{b}W^-$:
(a) Coulomb-gluon exchange between $t$ and $\bar{b}$,
(b) Coulomb-gluon exchange between $\bar{t}$ and ${b}$,
(c) Coulomb-gluon exchange between $b$ and $\bar{b}$, and
(d) transverse-gluon exchange between $b$ and $\bar{b}$.
}
\end{center}
\end{figure}

Corrections due to the final-state interactions (rescattering corrections)
that originate from the irreducible diagrams (a)--(d) in Fig.~\ref{fsifig}
are important particularly in the threshold region.
In fact their contributions are counted as 
${\cal O}(\alpha_s)={\cal O}(\beta )$ corrections to the leading
threshold enhancement \cite{r40}.
We calculate the
effect of each diagram on the $l^+$ energy-angular 
distribution.
We chose Coulomb-gauge for the QCD part in our calculations.

The contribution of diagram (a) 
(exchange of one Coulomb-gluon between $t$ and $\bar{b}$)
can be regarded, after integrating over the $\bar{b}W^-$ phase-space, as
a correction to the production process of $t$.
Thus, the production cross section 
$d\sigma_{\mbox{\scriptsize{Born}}}(e^+e^- \! \to t\bar{t})/d^3\vc{p}_t$
and the polarization vector 
$\mbox{\boldmath$\cal P$}_{\mbox{\scriptsize{Born}}}$
of the top quark
receive corrections by this diagram, whereas the decay distribution
$d\Gamma_{t \to bl^+\nu}(\mbox{\boldmath$\cal P$}_{\mbox{\scriptsize{Born}}})
/ dx_l d\Omega_l$
remains unaffected (except for the modification of 
$\mbox{\boldmath$\cal P$}_{\mbox{\scriptsize{Born}}}$).
In fact the contribution of this diagram can be incorporated by 
the following substitutions in Eq.\ (\ref{born}):
\bea
\frac{d\sigma_{\mbox{\scriptsize{Born}}}(e^+e^- \! \to t\bar{t})}
{d^3{\bf p}_t} 
\rightarrow
\frac{d\sigma_{\mbox{\scriptsize{Born}}}(e^+e^- \! \to t\bar{t})}
{d^3{\bf p}_t} 
\times
\left( 1 + \delta_a \right),
~~~~~
\mbox{\boldmath$\cal P$}_{\mbox{\scriptsize{Born}}}
\rightarrow
\mbox{\boldmath$\cal P$}_{\mbox{\scriptsize{Born}}}
+ \delta \mbox{\boldmath$\cal P$}_a
\label{diagrama1}
\eea
with
\bea
\delta_a &=& {\frac{1}{2}}
\left[ 
\psi_1(p,E) + \kappa C^0_\| \psi_{\rm _R}(p,E) \cos \theta_{te}
\right] ,
\label{diagrama2}
\\
\delta \mbox{\boldmath$\cal P$}_a &=&
{\frac{1}{2}} \left[ 1 - (C^0_\|)^2 \right]
\kappa \, \psi_{\rm _R}(p,E) \cos \theta_{te} \cdot \hat{\vc{n}}_\| .
\label{diagrama3}
\eea

The derivation of the formula goes as follows.
Since the relevant kinematical configuration lies in the soft gluon
region, we can use soft-gluon approximation
and factor out the part that depends on the loop-momentum $q^\mu$
(the propagators of the gluon, $\bar{b}$, $t$ and $\bar{t}$
together with the loop-integral)
outside the spinor structure, while the
remaining part is similar to the fully-differential
cross section of the Born-type diagram.
The latter part is factorized as before.
Due to the soft-gluon factor (the factor pulled outside), at this stage 
we may interpret that the 
production cross section and polarization vector of top quark
get corrections that depend on $t$ and $\bar{b}$ momenta.
Integrations over $dp_W^2$ and $d\phi_{bl}$ are the same as for the Born-type
diagram.
For integrations over $d\Phi_2(\bar{b}W^-)$, $dp_t^0/(2\pi )$ 
and $dq^0/(2\pi )$,
we follow the method described in Ref.\cite{r40}, Appendix D.
We are thus led to Eqs.\ (\ref{diagrama1})--(\ref{diagrama3}).

Diagram (b) (exchange of one Coulomb-gluon between $\bar{t}$ and $b$)
in Fig.\ \ref{fsifig} gives a correction that connects the production
and decay processes of the top quark. 
In fact one may incorporate the contribution of this diagram by
multiplying Eq.\ (\ref{born}) by a factor $[1+\xi (p,E,x_l,\cos\theta_{lt})]$,
where
\bea
\xi (p,E,x_l,\cos\theta_{lt})
= C_F \cdot 4\pi\alpha_s (\mu_B )
\int {\textstyle \frac{d^3\vc{q}}{(2\pi)^3} } \,
\frac{1}{|\vc{q}\! -\! \vc{p}_t|^3} \, \mbox{Re}
\left[ 
\frac{\tilde{G}^*(q,E)}{\tilde{G}^*(p,E)} 
\, {\displaystyle \int^{2\pi}_0} \!
{\textstyle \frac{d\phi_{bl}}{2\pi} } \,
\frac{|\vc{q}\! -\! \vc{p}_t|}
{\hat{\vc{n}}_b \! \cdot \! (\vc{q}\! -\! \vc{p}_t) + i\epsilon}
\right] .
\nonumber \\
\label{xi1}
\eea
$\hat{\vc{n}}_b$ denotes the unit vector in the direction of $b$.
After integration over $d\Omega_{\vc{q}}d\phi_{bl}$,
one may reduce the expression to a one-parameter integral form as
\bea
\xi (p,E,x_l,\cos\theta_{lt})
= C_F \cdot 4\pi\alpha_s (\mu_B )
\int^\infty_0 \! dq 
\left\{
w_{\rm _R} \,
\mbox{Re} \biggl[ \frac{\tilde{G}^*(q,E)}{\tilde{G}^*(p,E)} \biggl]
\, + \, w_{\rm _I} \,
\mbox{Im} \biggl[ \frac{\tilde{G}^*(q,E)}{\tilde{G}^*(p,E)} \biggl]
\right\}
\label{onepmint}
\eea
with
\bea
&&
w_{\rm _R} = \frac{1}{4\pi^2} \, \frac{q}{p^2-q^2}
\left\{
\theta ( z_+^2 \! -1 ) \cosh^{-1} |z_+| 
- \theta ( z_-^2 \! -1 ) \cosh^{-1} |z_-| 
\right\} ,
\\
&&
w_{\rm _I} = \frac{1}{4\pi^2}\, \frac{q}{|p^2-q^2|}
\left\{
\pi \theta \left(
( 1 + z_+ )( 1 + z_- )
\right) 
- \theta ( 1 - z_+^2 ) \cos^{-1} (z_+)
- \theta ( 1 - z_-^2 ) \cos^{-1} (z_-)
\right\} ,
\nonumber \\
\\
&&
z_\pm = \frac{q/p \pm \cos\theta_{Wl} \cos\theta_{lt}}{\sin\theta_{Wl} \sin\theta_{lt}},
~~~
\cos\theta_{Wl} =  \frac{1+y}{1-y} - \frac{2y}{x_l(1-y)},
~~~
\cos\theta_{lt} = \frac{\vc{p}_t\cdot\vc{p}_l}{|\vc{p}_t||\vc{p}_l|} .
\label{zangles}
\eea
Here, 
$\theta_{Wl}$ represents
the angle between $W^+$ and $l^+$ in the $t$ rest frame (given as
a function of $x_l$);
$\theta_{lt}$ represents the angle between $t$ and $l^+$ in 
the \ttbar c.m.\ frame\footnote{
Within our approximation, there is no distinction between $\cos\theta_{lt}$
and $- \cos\theta_{l\bar{t}}$, where $\theta_{l\bar{t}}$ denotes the
angle between $\bar{t}$ and $l^+$ 
{\it in the $t$ rest frame}.
}
($0 \le \theta_{Wl},\theta_{lt} \le \pi$).
The inverse functions
$\cosh^{-1}$ and $\cos^{-1}$ in the above formulas take their values within
$[0,\infty)$ and $[0,\pi]$, respectively.
$\theta (x)$ is the unit step function.
It is understood that the principal value should be taken
in the integration of the
$w_{\rm _R}$-term as $p \to q$.

We derived Eqs.\ (\ref{xi1}) and (\ref{onepmint}) in the following manner.
As in diagram (a), we used the soft-gluon approximation and factored out
a soft-gluon factor (loop-integral of the propagators of gluon, $b$, 
$t$ and $\bar{t}$).
The remaining part is same as the fully-differential cross section of 
the Born-type diagram except for the $t$ and $\bar{t}$ propagators
and Green's functions, which is again factorized.
This time, however, the correction cannot be interpreted as associated
with the top-quark production process since the soft-gluon factor
depends on the $b$ momentum.
The integrations over $dp_W^2$ and $d\Phi_2(\bar{b}W^-)$ are the same as those
for the  Born-type diagram.
The function $\xi$ given in Eq.(\ref{xi1}) is essentially the soft-gluon
factor integrated over 
$dp_t^0/(2\pi )$, $d^4q/(2\pi )^4$ and $d\phi_{bl}$.
Finally, to derive Eq.(\ref{onepmint}) from (\ref{xi1}), it is simpler to 
integrate over $d\Omega_{\vc{q}}$ before $d\phi_{bl}$.

Two noteworthy properties of $\xi (p,E,x_l,\cos\theta_{lt})$ are:
(1) its $l^+$-angular dependence enters only through $\cos\theta_{lt}$ and is
independent of the angle from the $e^-$ beam direction or from the top-quark
polarization vector, and
(2) it is purely determined by the QCD interaction and free of the
coupling parameters of electroweak interactions (except $y$).

As a non-trivial cross-check 
of the formula (\ref{onepmint}), we integrated 
$\xi \times \Gamma_t^{-1} d\Gamma_{t \to bl^+\nu}/ dx_l d\Omega_l$
over the
lepton energy-angular variables $\int dx_l d\Omega_l$ analytically
and reproduced
the one-parameter integral formula \cite{r40,r18} for the final-state
interaction correction (from diagram (b)) to the top-quark three-momentum 
distribution.

The contribution of diagram (c) 
(exchange of one Coulomb-gluon between $b$ and $\bar{b}$)
vanishes within our approximation. 
We show it in steps.
Using soft-gluon approximation, the contribution of this diagram to the
cross section can be written as
\bea
\frac{d\sigma_c}{d^3\vc{p}_tdx_ld\Omega_l} &=& \int
{\textstyle \frac{dp^0_t}{(2\pi)}} \, dp_W^2 \, d\phi_{bl} \, 
d\Phi_2(\bar{b}W^-) \,
\frac{1}{(2\pi)^3}\,  \frac{1}{(4\pi)^5} \times
T
\nonumber \\
&&
\times iC_F \cdot 4\pi\alpha_s \int
{\textstyle \frac{d^4k}{(2\pi)^4}} \,
\left[ D(p_t) \! +\! D(p_{\bar{t}}) \right]
\left[ D^*(p_t\! +\! k) \! +\! D^*(p_{\bar{t}}\! -\! k) \right]
\nonumber \\
&& ~~~~~~~~~~~~~~~~
\times 
\tilde{G}(p,E) \tilde{G}^*(|\vc{p}_t \! + \! \vc{k}|,E) \,
\frac{1}{k^0 - \hat{\vc{n}}_b\! \cdot\! \vc{k} - i\epsilon} \cdot
\frac{1}{-k^0+\hat{\vc{n}}_{\bar{b}}\! \cdot \! \vc{k} - i\epsilon} \cdot
\frac{1}{|\vc{k}|^2}
\nonumber \\
&& 
+ \mbox{complex conj.} ,
\label{diagramc}
\eea
where $k^\mu$ is the gluon momentum and
\bea
D(p_t) = \frac{1}{p_t^0 -m_t - \vc{p}_t^2/2m_t + i\width /2}
\eea
denotes the non-relativistic top-quark propagator.
$T$ is the contraction of hadronic and leptonic tensors resulting from
the spinor traces after a soft-gluon factor (the second and third
lines) is taken out.
It coincides with the fully-differential cross section of 
the Born-type diagram except for the $t$ and $\bar{t}$ propagators
and Green's functions; hence $T$ is real.
Next we integrate over $dp_t^0/(2\pi)$, $dk^0/(2\pi)$ and $d\Omega_{\vc{k}}$.
Let us define
\bea
&&
I(\vc{k},\hat{\vc{n}}_b , \hat{\vc{n}}_{\bar{b}})  
\nonumber \\
&&\equiv ~
\int {\textstyle \frac{dp^0_t}{(2\pi)}\frac{dk^0}{(2\pi)}}
\left[ D(p_t) \! +\! D(p_{\bar{t}}) \right]
\left[ D^*(p_t\! +\! k) \! +\! D^*(p_{\bar{t}}\! -\! k) \right]
\frac{1}{k^0 - \hat{\vc{n}}_b\! \cdot \! \vc{k} - i\epsilon} \cdot
\frac{1}{-k^0+\hat{\vc{n}}_{\bar{b}}\! \cdot \! \vc{k} - i\epsilon}
\nonumber \\
&&\simeq ~
\frac{1}
{( \hat{\vc{n}}_b -\hat{\vc{n}}_{\bar{b}})\! \cdot \! \vc{k} + i \epsilon} \,
\biggl(
\frac{1}{\hat{\vc{n}}_b\! \cdot\! \vc{k} + i\width } +
\frac{1}{- \hat{\vc{n}}_{\bar{b}}\! \cdot\! \vc{k} + i\width }
\biggl) .
\eea
Noting that the contribution of diagram (c) to the cross section
comes solely from the gluon momentum region
\bea
k^0, |\vc{k}| \sim \alpha_s^2 m_t ,
\label{kinregion}
\eea
we keep $|\vc{k}|$ within this region.
Then we may substitute 
$\tilde{G}^*(|\vc{p}_t+\vc{k}|,E) \to \tilde{G}^*(p,E)$
in Eq.\ (\ref{diagramc}) since the difference is higher order, and find
\bea
\frac{d\sigma_c}{d^3\vc{p}_tdx_ld\Omega_l} &=& \int
dp_W^2 \, d\phi_{bl} \, 
d\Phi_2(\bar{b}W^-) \, \frac{1}{(2\pi)^6}\,  \frac{1}{(4\pi)^5} 
\times T
\nonumber \\
&&
\times iC_F \cdot 4\pi\alpha_s \left| \nrg \right|^2 \times
{\hbox to 18pt{
\hbox to -3pt{$\displaystyle \int$} 
\raise-15pt\hbox to 7pt{$\scriptstyle |\vc{k}|\sim \alpha_s^2 m_t$} 
}}
|\vc{k}|^2 d |\vc{k}| \times \frac{1}{|\vc{k}|^2}
\int d\Omega_{\vc{k}} \, 
I(\vc{k},\hat{\vc{n}}_b , \hat{\vc{n}}_{\bar{b}})  
\nonumber \\
&+& \mbox{complex conj.}
\eea
It is easy to see that $\int d\Omega_{\vc{k}} \, 
I(\vc{k},\hat{\vc{n}}_b , \hat{\vc{n}}_{\bar{b}})$
is real using the symmetry of $d\Omega_{\vc{k}}$ 
under $\vc{k} \to -\vc{k}$.
Thus, we conclude $d\sigma_c/d^3\vc{p}_tdx_ld\Omega_l = 0$ in our
approximation.

In fact the same proof can be applied to show quite generally that 
the contribution of diagram (c) vanishes at \oalfs
{\it provided} one calculates a cross section where the top-quark energy is 
integrated out; for example, the top-quark three-momentum distribution.  
This is no longer the case when one considers a cross section that depends 
explicitly on the top-quark energy; for example, the top-quark four-momentum 
distribution. Then the diagram in question {\it does} contribute.

One can also show in a similar way that the kinematical regions
Eq.\ (\ref{kinregion}) in diagrams (a) and (b) do not contribute to
the cross section $d\sigma/d^3\vc{p}_t dx_l d\Omega_l$ and that
only the gluon momentum region where $|\vc{k}| \sim \alpha_s m_t \gg \width$ 
is relevant.
We took advantage of this fact in deriving 
Eqs.\ (\ref{diagrama1})--(\ref{onepmint}).

\begin{figure}
\begin{center}
  \leavevmode
  \epsfxsize=16.cm
  \epsffile{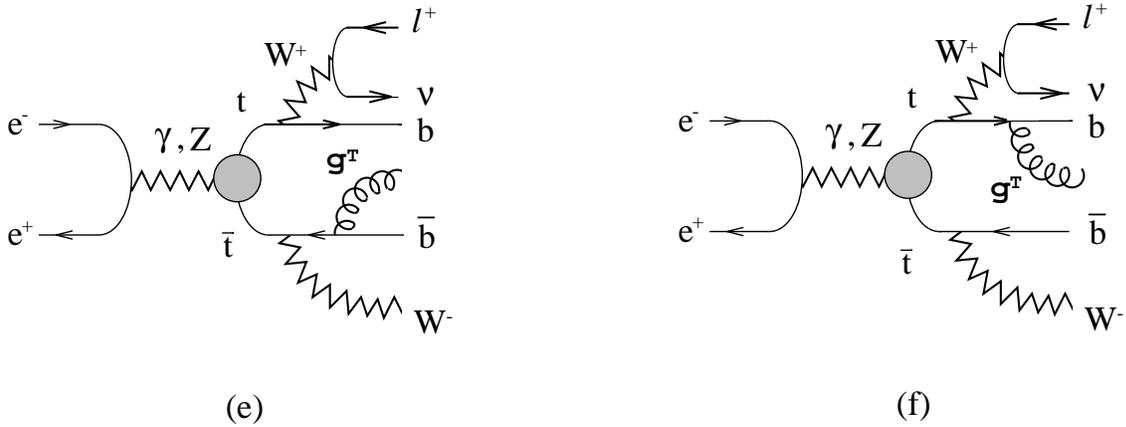}\\
  \caption[]{
        \label{gemissionfig}
Diagrams for real-gluon emission process
$e^+e^- \to t\bar{t} \to bl^+\nu\bar{b}W^- g$.
}
\end{center}
\end{figure}

It can be shown using similar techniques that the contribution of diagram (d) 
(exchange of one transverse gluon between $b$ and $\bar{b}$)
gets canceled when it is added 
to that of the corresponding real-gluon emission diagram
(interference of diagrams (e) and (f) in Fig.~\ref{gemissionfig}).
This cancellation is consistent with the same cancellation that was
found in the calculation of the top-quark momentum distribution \cite{r40}.
The contribution from each of these diagrams comes from the gluon momentum
region $|\vc{k}| \simlt \alpha_s^2 m_t$ and
is in fact logarithmically divergent due to a soft-gluon singularity.

The cancellations of the final-state interaction corrections
at the various levels of inclusive cross sections are summarized
in section 7.
\medbreak

Now let us compare our formulas and the argument given in the previous
section.
In subsections 3.b and 3.c, the parity-violating nature of the
electroweak interactions in top production and decay
played an essential role, while
this was not the case in subsections 3.a and 3.d.
We find that correspondingly the
$\cos\theta_{te}$ term of $\delta_a$
and $\delta \mbox{\boldmath$\cal P$}_a$ 
contain electroweak coupling parameters
(through $C^0_\|$), while the symmetric term of $\delta_a$ and
$\xi$ are independent of these electroweak parameters.
More precisely, the $\cos\theta_{te}$ term of $\delta_a$
and $\delta \mbox{\boldmath$\cal P$}_a$ 
have the forms anticipated in subsections 3.b and 3.c, respectively, 
if the function $\psi_{\rm _R}(p,E)$ is positive.
Indeed, the numerical evaluation in Ref.\ \cite{r26} shows that 
$\psi_{\rm _R}(p,E) \simgt 0$ holds in the entire threshold region.
Besides, an additional coefficient $\left[ 1 - (C^0_\|)^2 \right]$
in $\delta \mbox{\boldmath$\cal P$}_a$ can be understood 
within our previous argument
in the extreme cases $C^0_\| = \pm 1$.
Namely, if the top quark is 100\% polarized, there will be no
contamination from the opposite spin so that the correction should
disappear.
It may be interesting to note that the final-state interaction
corrections to the polarization vector vanishes for the ideally polarized
top quarks, $C^0_\| = \pm 1$.

\subsection{Formula Including Full \oalfs Corrections}
In summary, the energy-angular distribution of $l^+$ including full
${\cal O}(\alpha_s) = {\cal O}(\beta)$ corrections can be cast into a form
\bea
\frac{d\sigma(\epem \! \to \ttbar \to bl^+\nu\bar{b}W^-)}
{d^3\vc{p}_t dx_l d\Omega_l}
=
\frac{d\sigma(e^+e^- \! \to t\bar{t})}
{d^3\vc{p}_t} \times
\frac{1}{\Gamma_t} \,
\frac{d\Gamma_{t \to bl^+\nu}
(\mbox{\boldmath$\cal P$}) }
{dx_l d\Omega_l}  \times \left( 1 + \xi \right)
\label{fullformula}
\eea
with
\bea
&&
\frac{d\sigma(e^+e^- \! \to t\bar{t})}
{d^3\vc{p}_t} 
= \frac{d\sigma^0_{t\bar{t}}}
{d^3\vc{p}_t} 
\times
\left[
1 + \frac{1}{2}\psi_1 + 
( 2 C_{FB} \varphi_{\rm _R} + \frac{\kappa}{2}C_\|^0\psi_{\rm _R} )
\cos \theta_{te}
\right] ,
\label{prcs}
\\
&&
\mbox{\boldmath$\cal P$}
=
\mbox{\boldmath$\cal P$}_{\mbox{\scriptsize{Born}}}
+ \delta \mbox{\boldmath$\cal P$}_a .
\label{modpolvec}
\eea
The above distribution is obtained as the sum of the cross sections for
$e^+e^- \to t\bar{t} \to bl^+\nu \bar{b}W^-$ 
and
$e^+e^- \to t\bar{t} \to bl^+\nu \bar{b}W^-g$. 
An independent emission of a gluon from the $t$ or $\bar{t}$ side has been
included in Eq.\ (\ref{born}), while the interference of both
has been incorporated in conjunction with the final-state interaction 
diagram (d) in subsection 4.b.

Experimentally the top-quark four-momentum $p_t^\mu$
will necessarily be reconstructed from the
$\bar{b}W^-$ system in the study of the
$l^+$ distribution.
For the case with a gluon in the final state, we assign the 
``top-quark momentum'' as
\bea
\begin{array}{llll}
\mbox{Case (A)}: & p_t \equiv (p_{e^-}+p_{e^+})
- (p_{\bar{b}} + p_{W^-} + p_g)
& \mbox{if} &
(p_{\bar{b}} + p_{W^-} + p_g)^2 - m_t^2 \simlt m_t\width  ,
\\
\mbox{Case (B)}: & p_t \equiv (p_{e^-}+p_{e^+})
- (p_{\bar{b}} + p_{W^-}) &
\mbox{if} & (p_{\bar{b}} + p_{W^-})^2 - m_t^2 \simlt m_t\width .
\end{array}
\nonumber
\eea
(See, however, the discussion in section 7.)
In other kinematical configurations the cross section is suppressed.
Experimentally there will be a corresponding cut in the $\bar{b}W^-$
invariant mass.
If both conditions in 
(A) and (B) are satisfied simultaneously, the gluon should
necessarily be soft, and there will be no difference within our
approximation between the cross 
sections corresponding to the above two assignments of the top-quark
momentum.

One comment is in order here.
In defining the ``rest frame'' of a top-quark in the study
of $l^+$ distribution, one may
use either $p_t^\mu$ or $\tilde{p}_t^\mu$ (defined in Eq.(\ref{replpt})).
The difference of the cross sections based on the two definitions
is of ${\cal O}(\alpha_s^2)$ which is beyond the scope of
our approximations.
Thus, the cross sections defined in both definitions should be measured
in experiment and compared.
It will serve as a cross-check for the stability of our prediction.

\section{Numerical Results}
\cleqn
\clfn

In this section 
we examine the effects of the final-state interaction
corrections on the $l^+$ energy-angular distribution numerically,
and compare the results with the qualitative argument given in
section 3. 
The numerical results are obtained
using both the coordinate-space approach developed in Refs.\ \cite{r8,r9}
and the momentum-space approach developed in Refs.\ \cite{r35,r36}.
Conventionally these two approaches have been used independently by different 
groups, and
this is the first time to make a direct comparison of the cross sections
calculated in both approaches.
Some of the produced results are slightly different.
We set $m_t=175$~GeV, $\alpha_s(M_Z)=0.118$, $P_{e^+}=0$ and
$\alpha=1/128$ in all our analyses.

We first examine the contribution of diagram (a) in Figs.~\ref{fsifig}
(Coulomb interaction between $t$ and $\bar{b}$) as given in
Eqs.\ (\ref{diagrama1})--(\ref{diagrama3}).
\begin{figure}
\begin{center}
  \leavevmode
  \epsfxsize 80mm  \epsffile{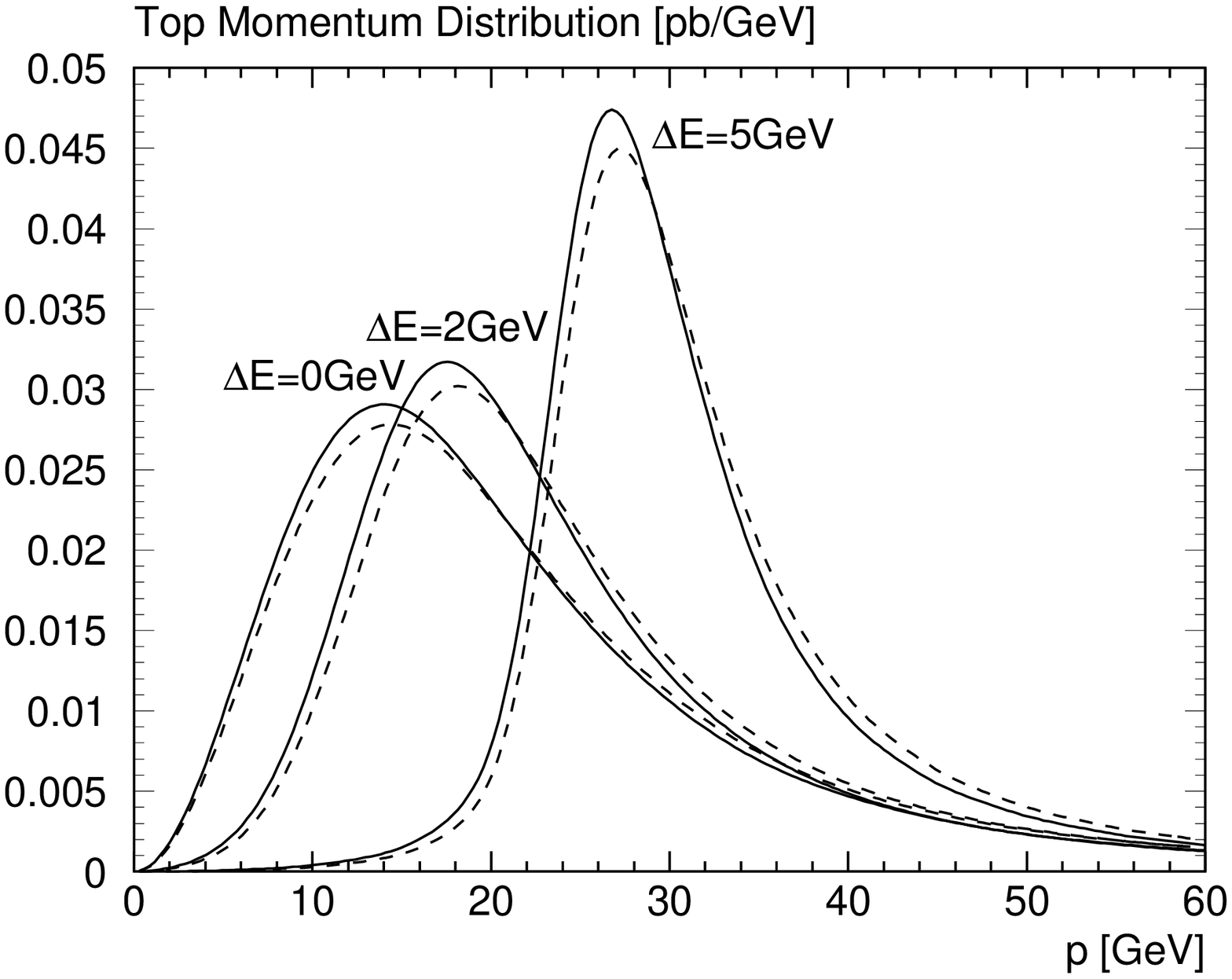}
  \epsfxsize 80mm \epsffile{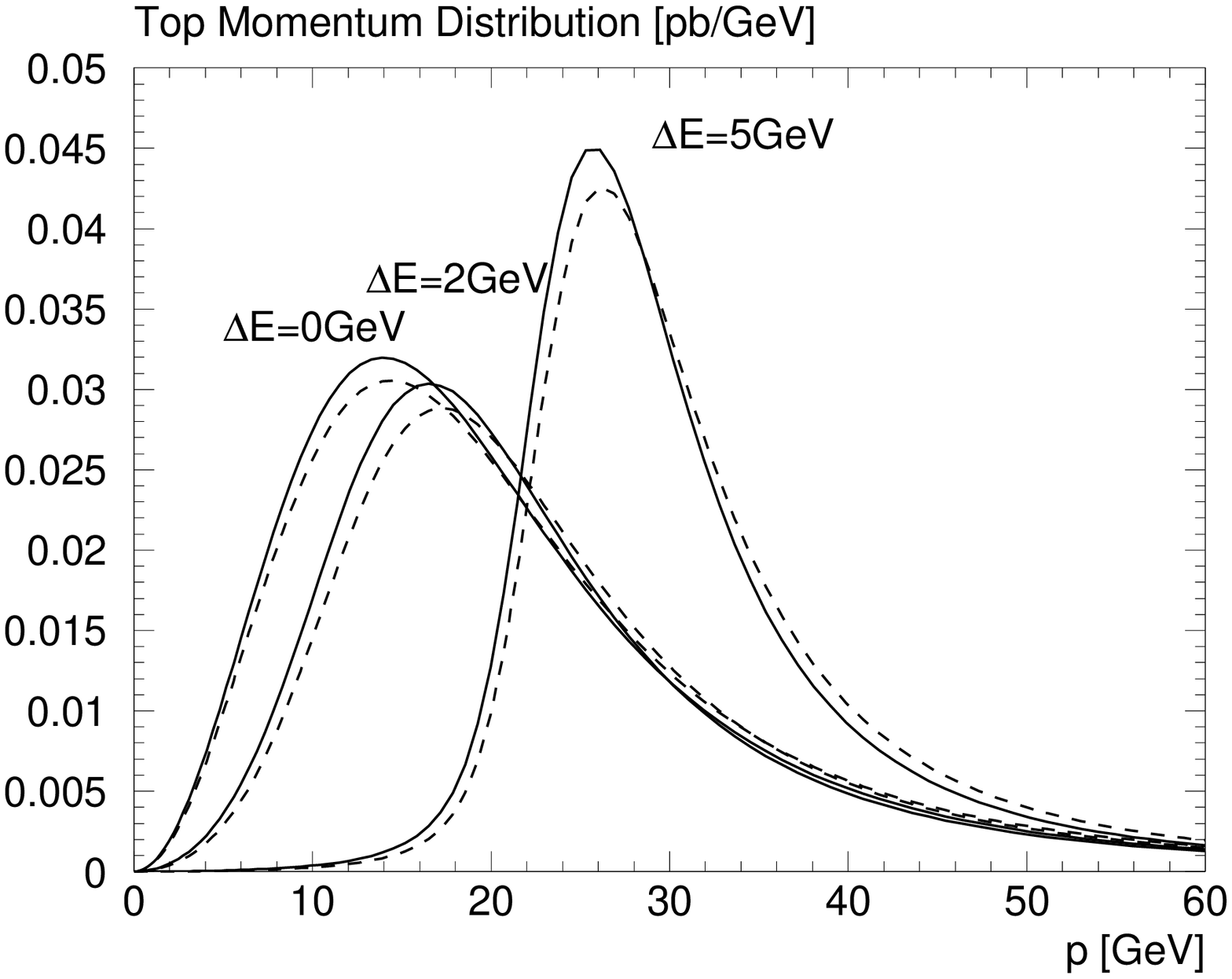}
  \end{center}
  \caption[]{\label{momdist}Top-quark momentum distributions
    $d\sigma/d|\vc{p}_t|$ obtained from 
    Eq.\ (\ref{diagrama1}) for various c.m.\ energies measured from
   the lowest-lying resonance, $\Delta E = \sqrt{s} - M_{1S}$.
   Solid lines and dashed lines represent the distributions with
   and without the final-state interaction between $t$ and $\bar{b}$
   ($\delta_a$), respectively.  We set $P_{e^-}=0$. The left figure
   corresponds to the coordinate-space calculation, the right one to
   the momentum-space calculation.}
\end{figure}
Shown in Fig.\ \ref{momdist} are the top-quark 
momentum distribution $d\sigma / d|\vc{p}_t|$
for various c.m.\ energies measured 
from the lowest-lying resonance,\footnote{
$M_{1S}$ is defined as the
real part of the position of the lowest-lying
resonance pole in the complex energy
plane.
The energy measured from $M_{1S}$ is more convenient than the energy measured
from the threshold ($E=\sqrt{s}-2m_t$) when we compare different potentials
in the literature.
} 
$\Delta E = \sqrt{s} - M_{1S}$.
This is calculated from the
top-quark production cross section Eq.\ (\ref{diagrama1}).
As expected, the top-quark momentum is reduced.
The effects are half in magnitude as compared to the
final-state interaction corrections given in Refs.\ \cite{r40,r26}
since only the interaction between $t$ and $\bar{b}$ is included here.

\begin{figure}
\begin{center}
  \epsfxsize 80mm \mbox{\epsffile{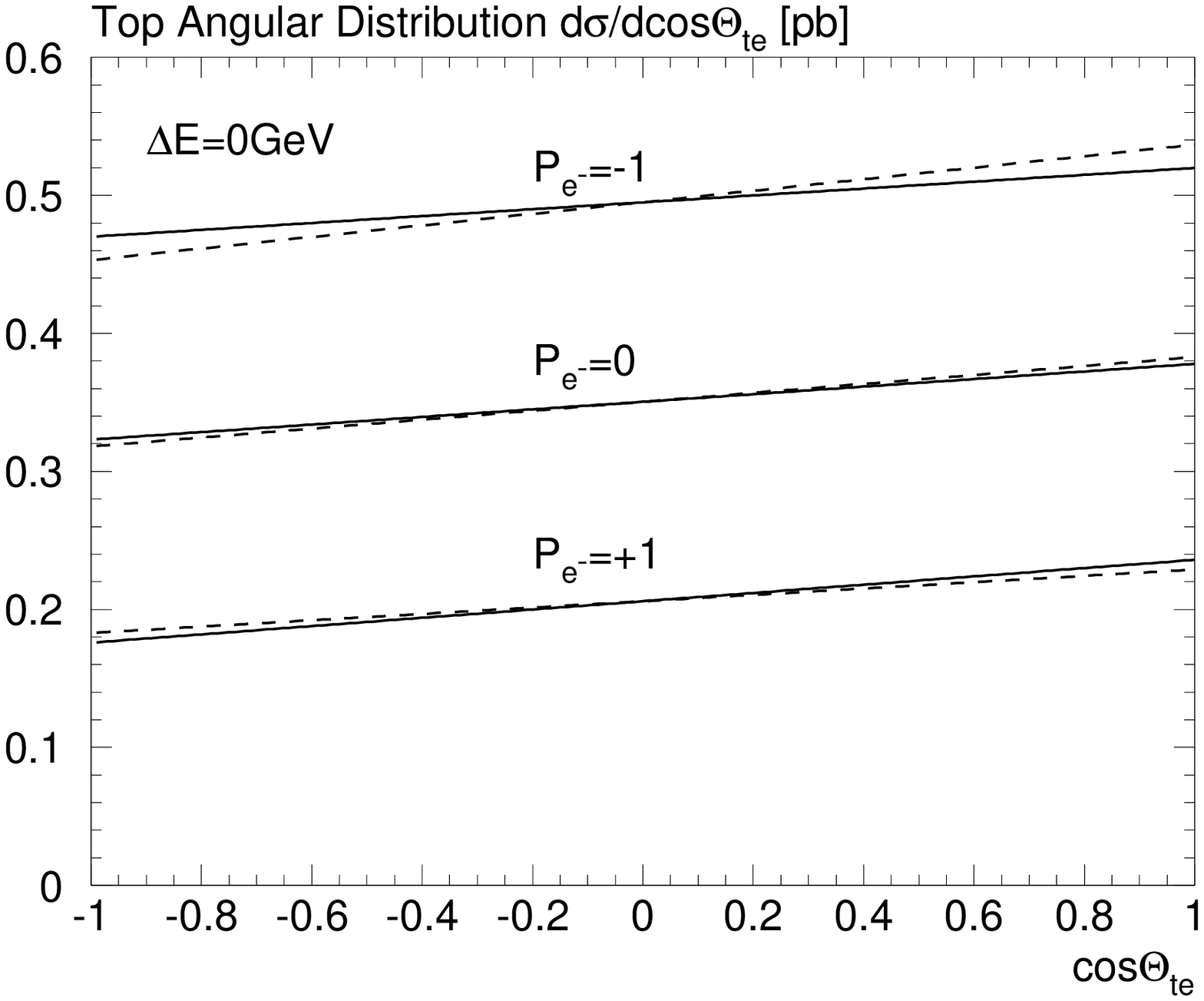}} 
  \epsfxsize 80mm \mbox{\epsffile{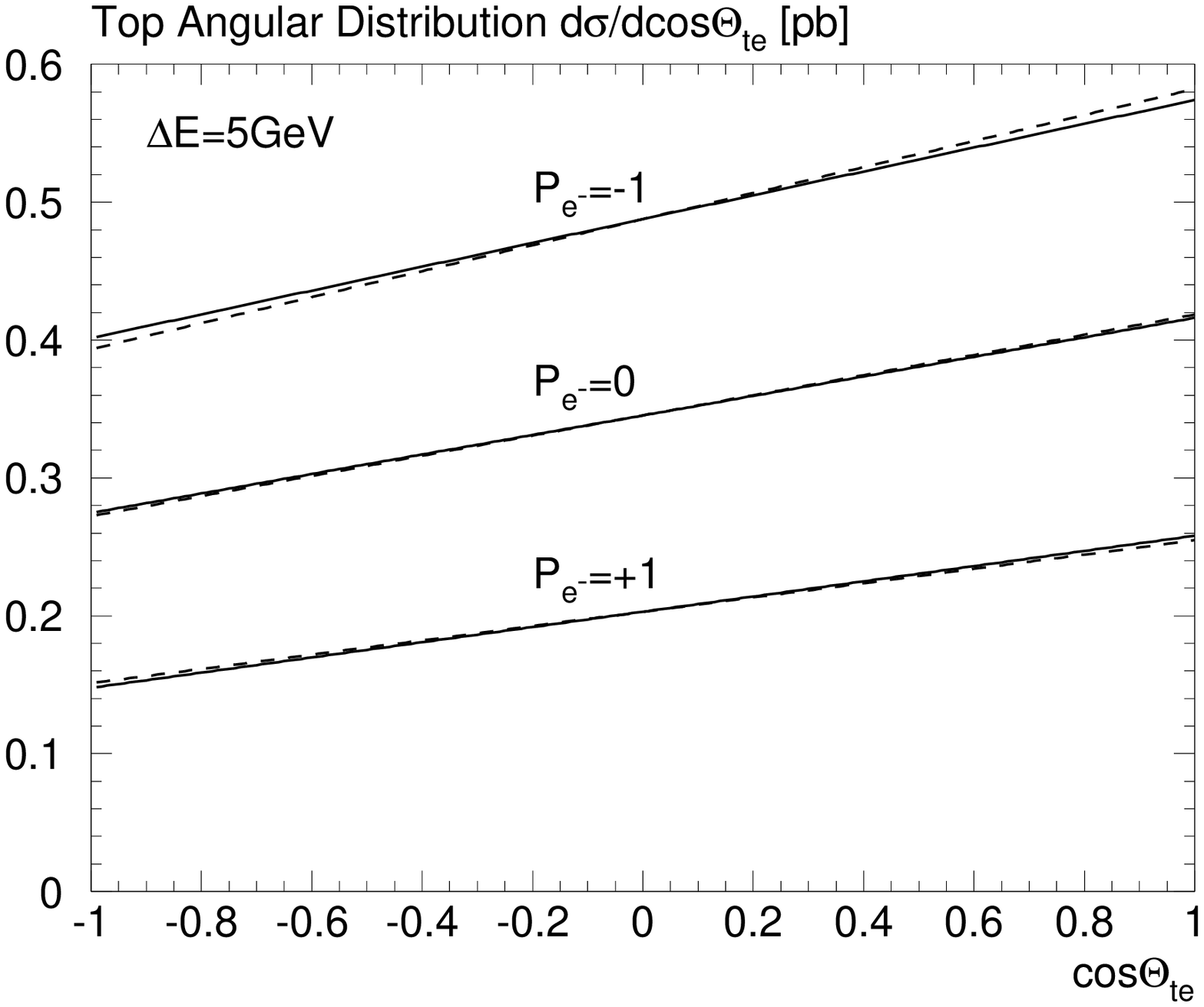}} \\
  a) Coordinate-space approach results \\
  \epsfxsize 80mm \mbox{\epsffile{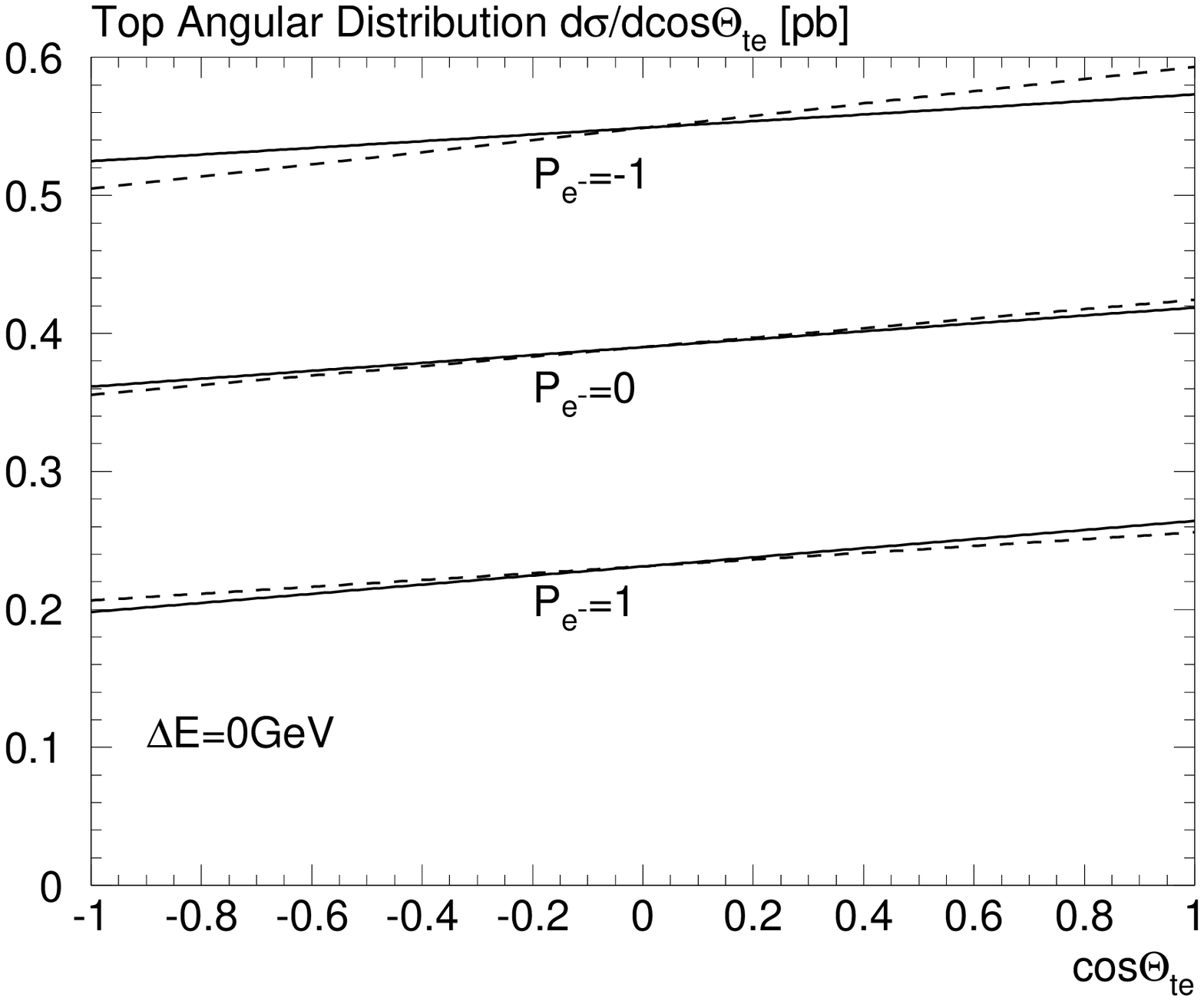}} 
  \epsfxsize 80mm \mbox{\epsffile{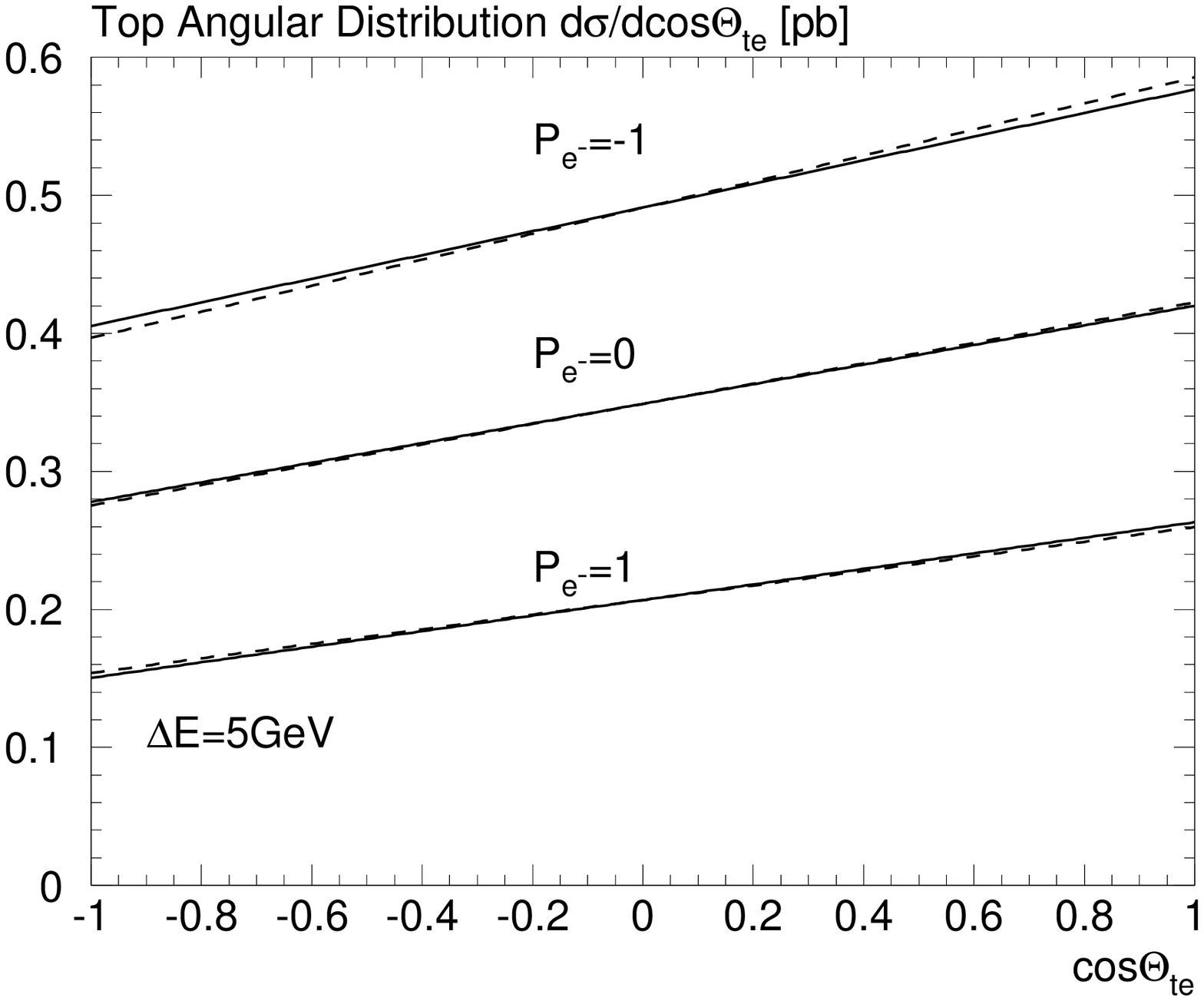}} \\
  b) Momentum-space approach results
\end{center}
\caption{\label{angdist}
Top-quark angular distribution $d\sigma / d\cos\theta_{te}$
obtained from 
Eq.\ (\ref{diagrama1}) for various electron polarizations $P_{e^-}$.
Solid lines and dashed lines represent the distributions with
and without the final-state interaction between $t$ and $\bar{b}$
($\delta_a$), respectively.  
}
\end{figure}

\begin{figure}
\begin{center}
  \epsfxsize 80mm \mbox{\epsfbox{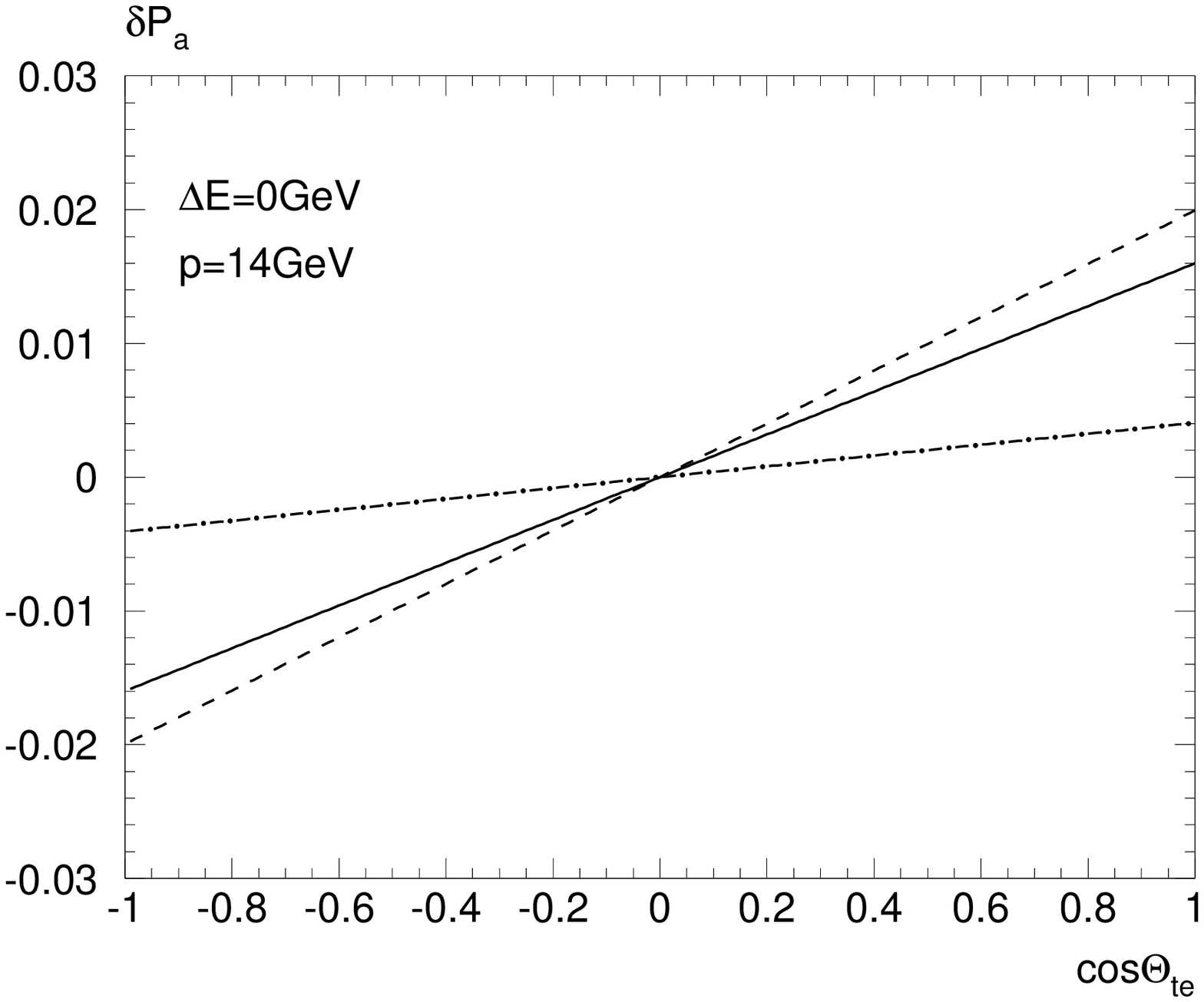}} 
  \epsfxsize 80mm \mbox{\epsfbox{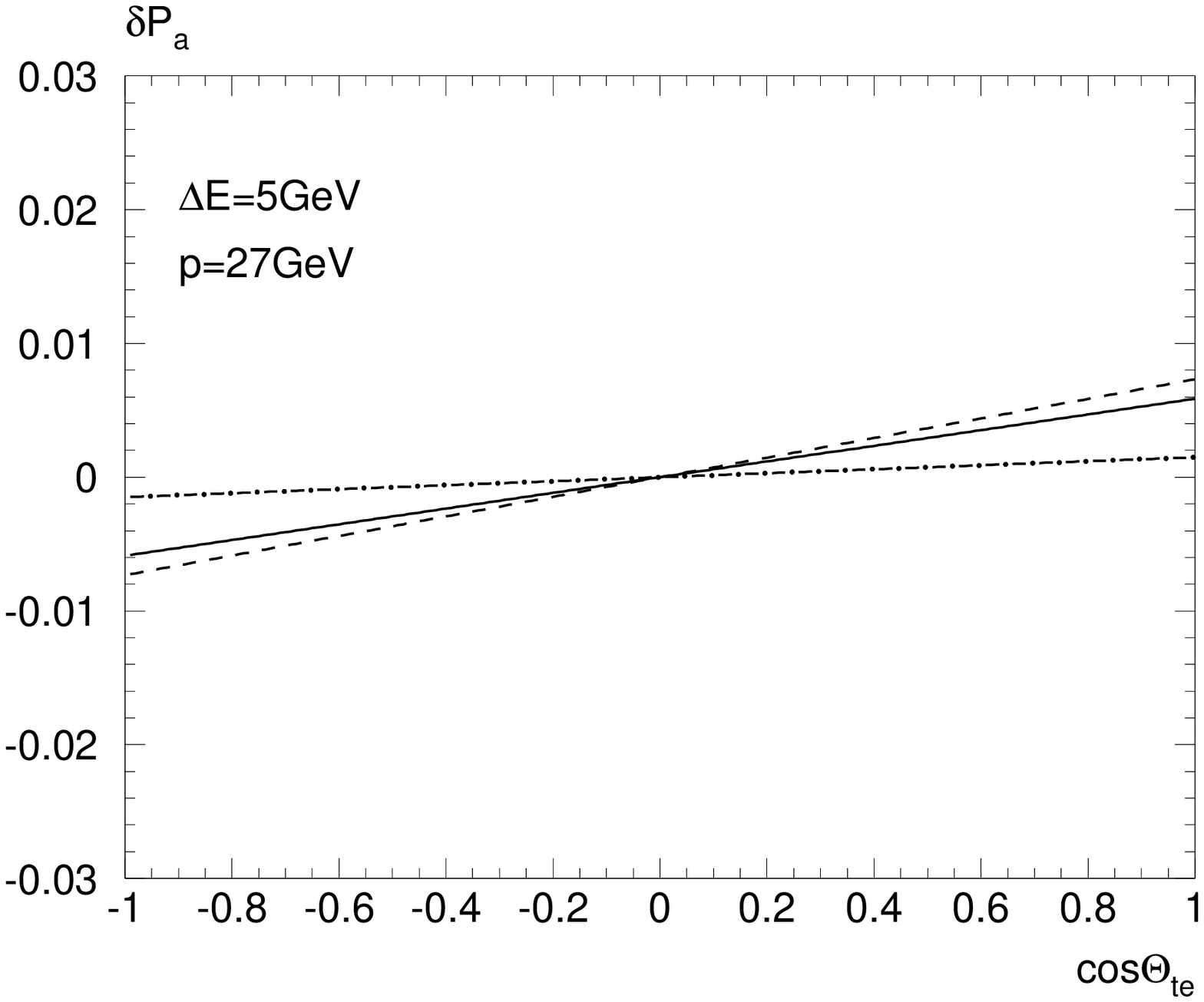}} \\
  a) Coordinate-space approach results \\
  \epsfxsize 80mm \mbox{\epsfbox{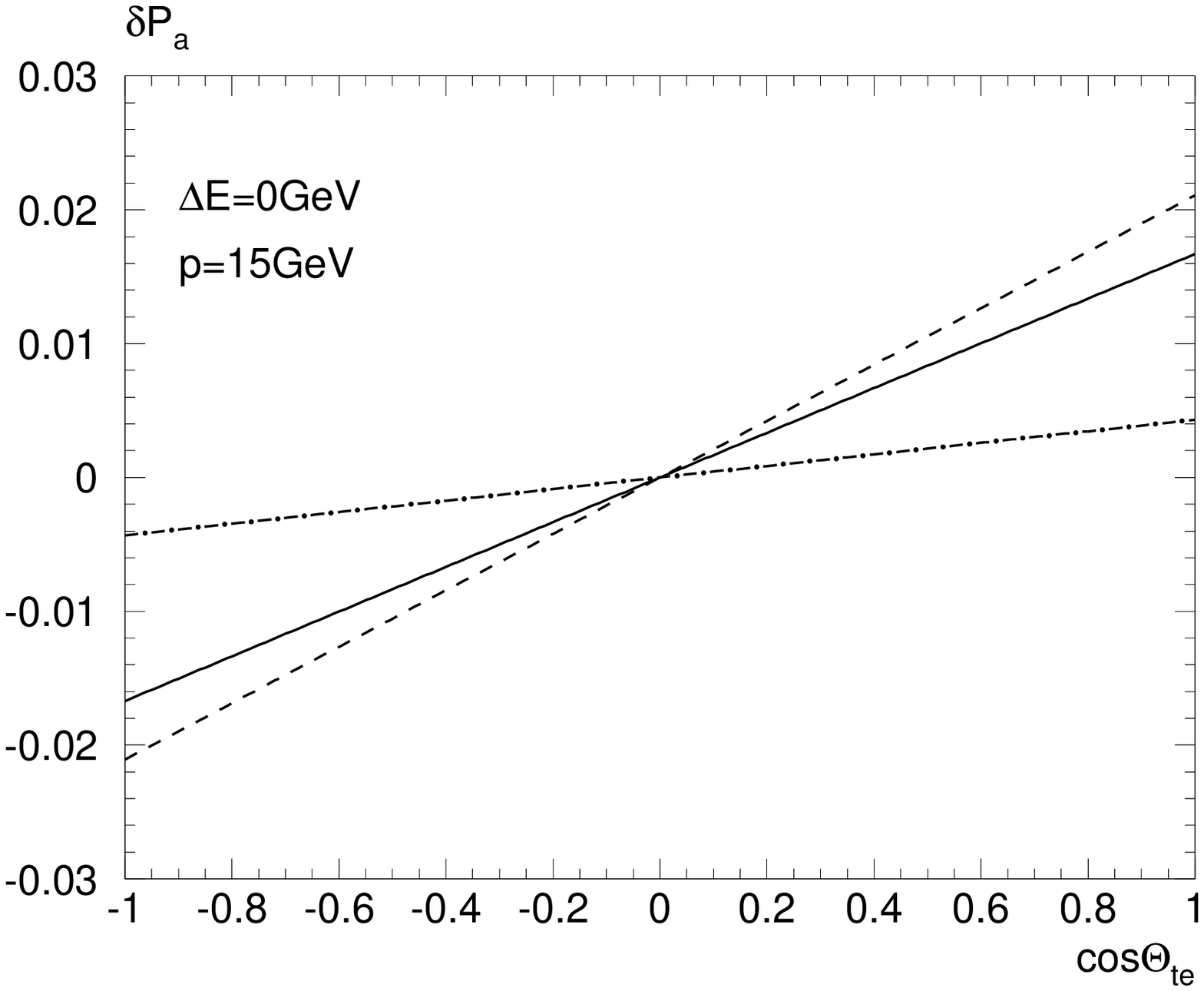}}
  \epsfxsize 80mm \mbox{\epsfbox{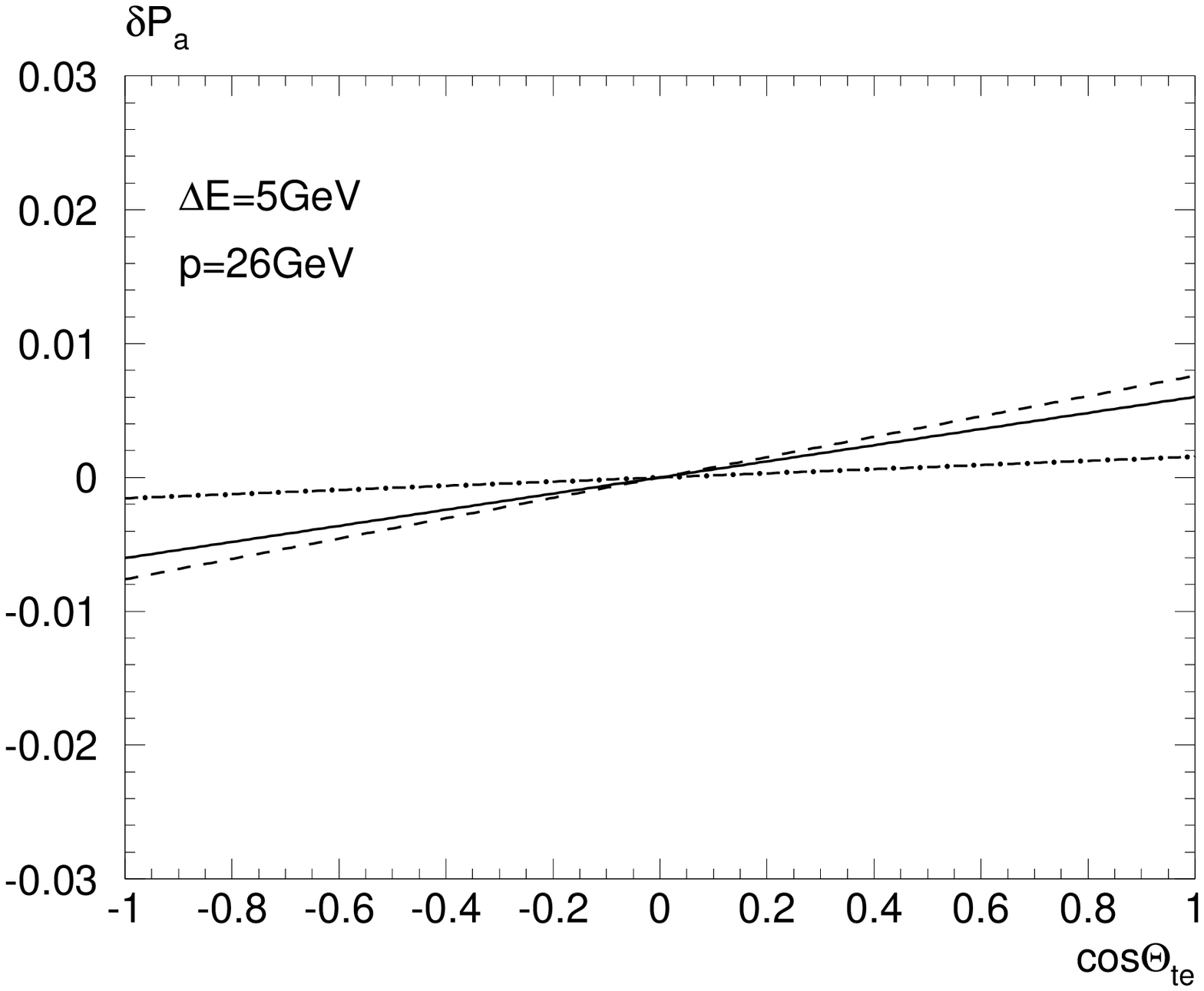}} \\
  b) Momentum-space approach results
\end{center}
\caption{\label{delP}
Final-state interaction correction to the $\hat{\vc{n}}_\|$ component
of the top-quark polarization vector, $\delta \mbox{\boldmath$\cal P$}_a$, for
$P_{e^-}=+ 0.8$ (solid line), 0 (dash line), $-0.8$ (dot-dash line).
}
\end{figure}

We show the angular distribution of the top-quarks in Figs.~\ref{angdist}.
It can be seen that the final-state interaction increases the top-quark 
distribution in the
forward direction for $P_{e^-}=1$.
This is consistent with our argument in subsection 3.b since
in leading-order approximation
$t$ and $\bar{t}$ have their
spins aligned perfectly in the $\hat{\vc{n}}_\|$
direction for this $e^-$ polarization. 
Oppositely we see that the final-state interaction decreases the top-quark
distribution in the forward direction for $P_{e^-}=-1$.
We note that the top quark has a natural
polarization 
$\mbox{\boldmath$\cal P$} \simeq -0.4 \hat{\vc{n}}_\|$
for unpolarized $e^+e^-$ beams. 
Hence, the sign of the correction
is the same as in the $P_{e^-}=-1$ case.
Also we show corrections to the top-quark polarization vector
in Fig.~\ref{delP}.
Although the qualitative behaviour meets our expectation, the magnitude of
the correction is rather small.
Note that $\delta \mbox{\boldmath$\cal P$}_a$ vanishes for $P_{e^-}=\pm 1$
since $C^0_\|=\pm 1$.

\begin{figure}
\begin{center}
  \epsfxsize=80mm  \mbox{\epsfbox{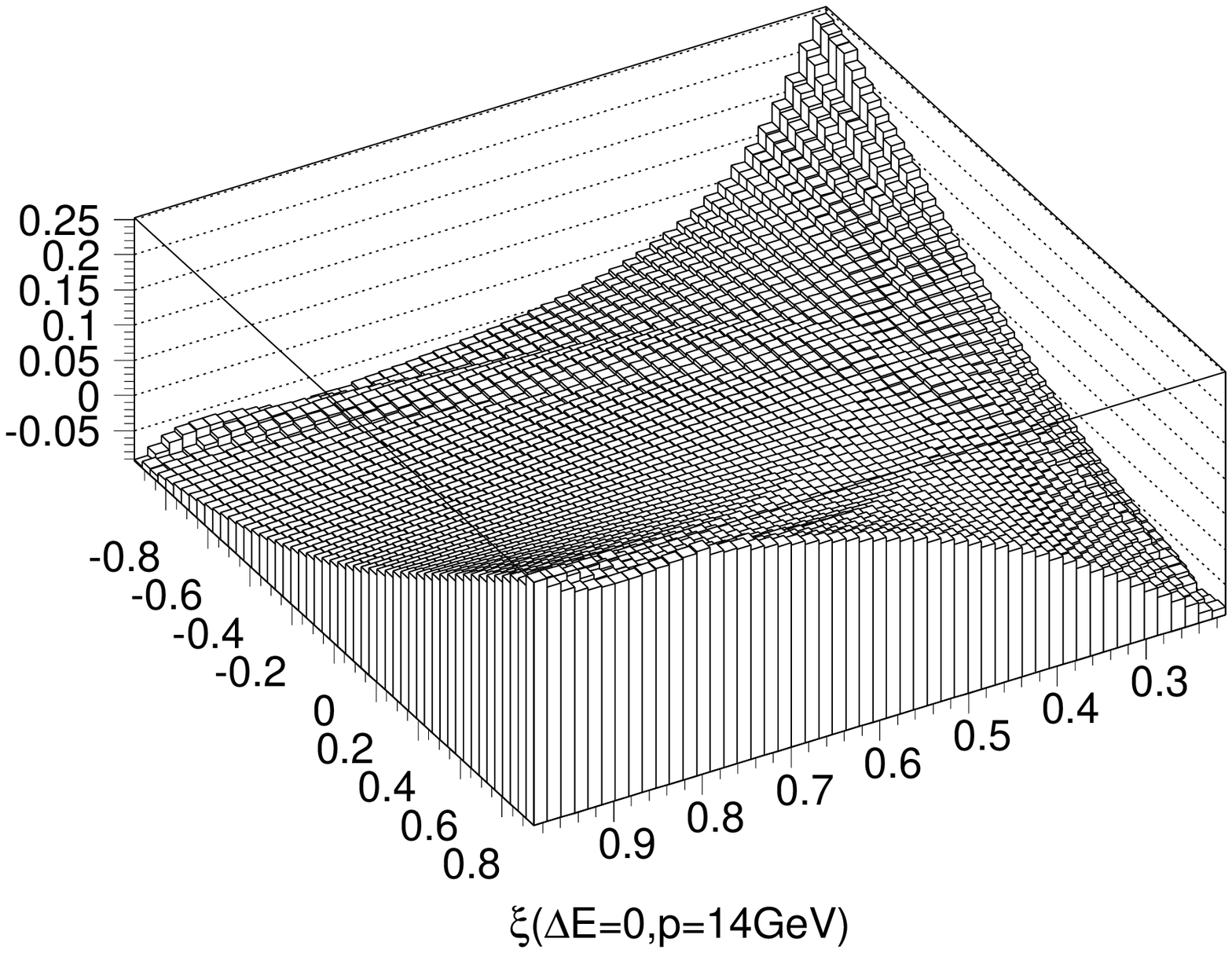}}
  \epsfxsize=80mm  \mbox{\epsfbox{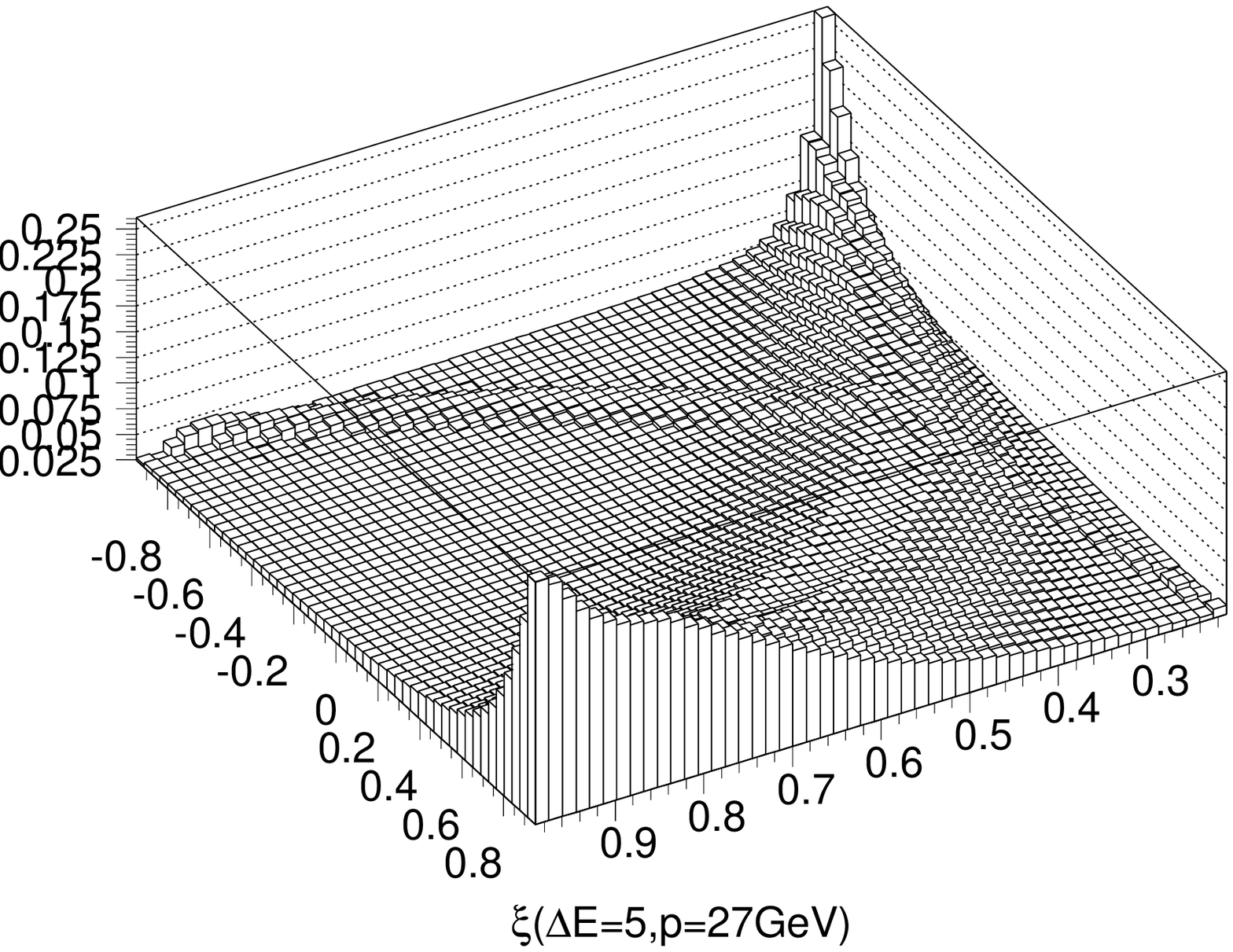}} \\
  a) Coordinate-space approach results \\  
  \epsfxsize=80mm  \mbox{\epsfbox{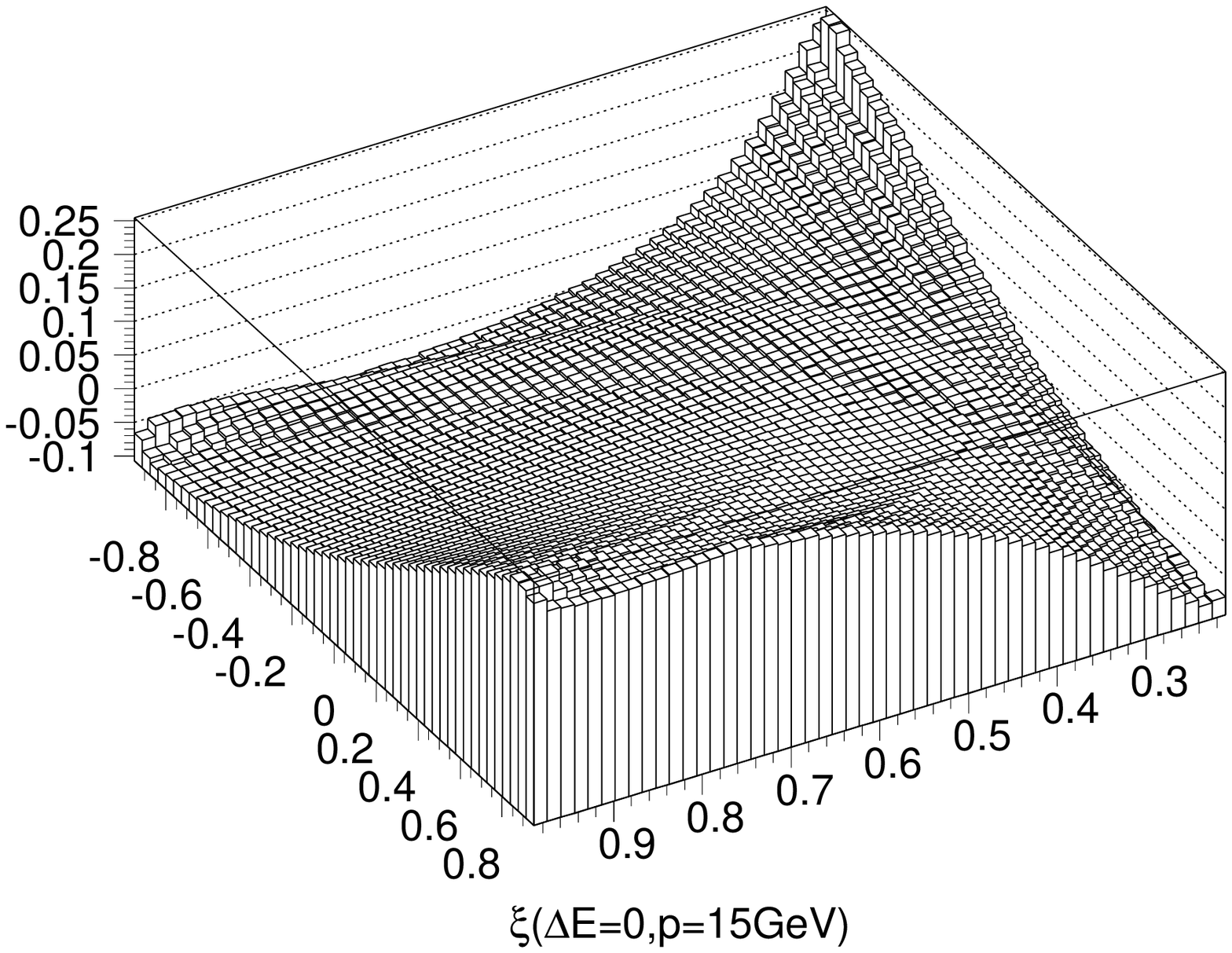}}
  \epsfxsize=80mm  \mbox{\epsfbox{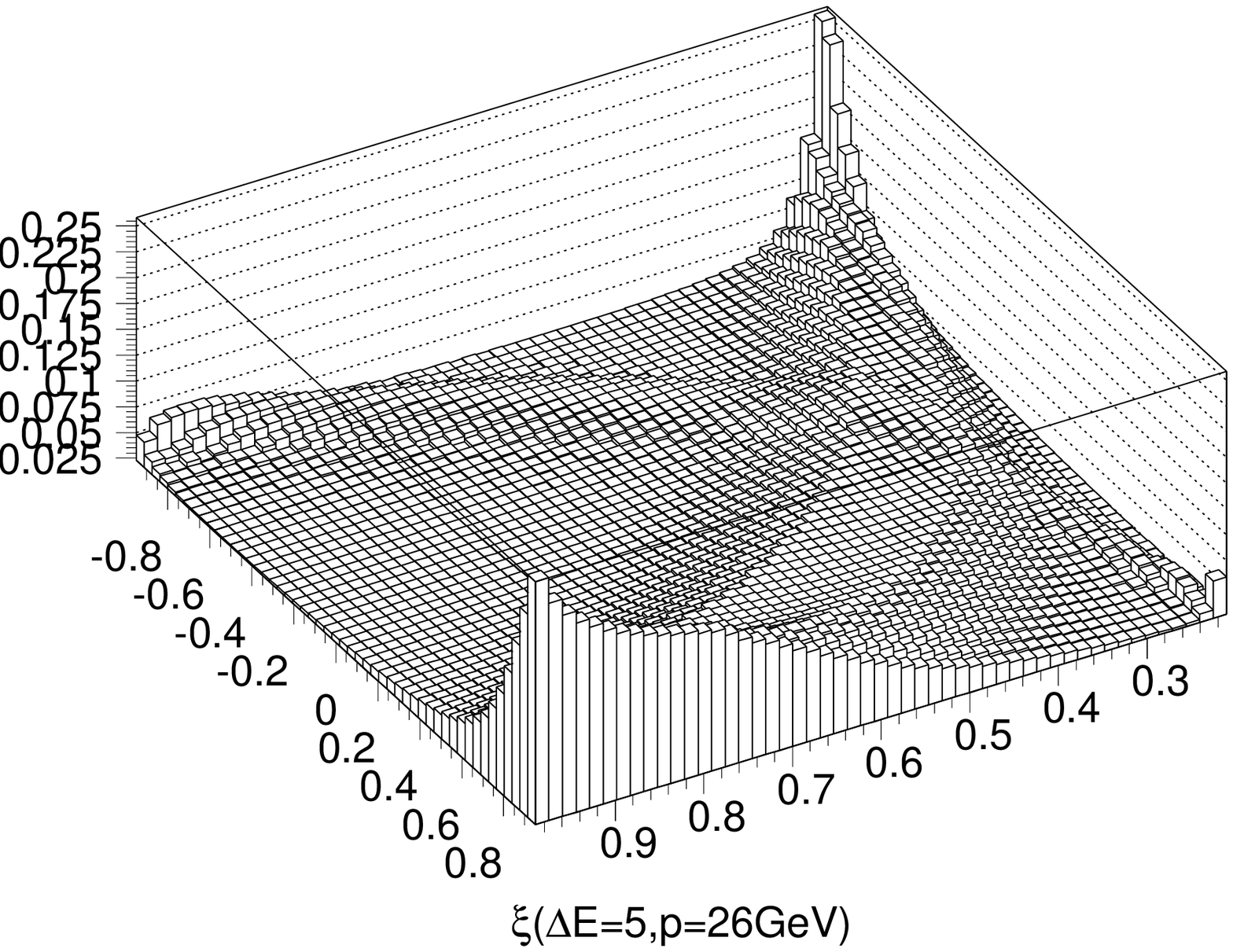}} \\
  b) Momentum-space approach results
\end{center}
  \caption[]{
        \label{xi}
Three-dimensional plots of $\xi$ as a function of $x_l$ ($x$-axis) and
$\cos\theta_{te}$ ($y$-axis) at the peak momenta of the 
$|\vc{p}_t|$-distribution in Fig.~\ref{momdist}:
$\Delta E=0$~GeV and $\Delta E=5$~GeV.
}
\end{figure}

Next we investigate the irreducible (non-factorizable) correction
that stems from the Coulomb interaction between $\bar{t}$ and $b$.
The correction factor $\xi(p,E,x_l,\cos \theta_{lt})$ given
in Eqs.\ (\ref{onepmint})--(\ref{zangles})
depends on four parameters, two of which 
specify the lepton configuration: $x_l$ and $\cos \theta_{lt}$.
Therefore, we will examine the dependence of $\xi$ on these 
two parameters
for several $(p, \Delta E)$ combinations.
We fix the top-quark momentum $p$ to be the peak momentum
of the distribution for
each $\Delta E$, hence its values are slightly different for the
two numerical approaches;
see the top-quark momentum distributions in Fig.~\ref{momdist}.
Shown in Fig.~\ref{xi} are 3-dimensional
plots of $\xi$ as a function of $x_l$ and $\cos \theta_{lt}$.
One can see that in the figures $\xi$ takes comparatively large
positive values for
either ``small $x_l$ and $\cos \theta_{lt} \simeq -1$'' or
``large $x_l$ and $\cos \theta_{lt} \simeq +1$''.
Oppositely,
in the other two corners of the $x_l$--$\cos \theta_{lt}$ plane
$\xi$ becomes small or becomes negative for smaller $\Delta E$.
The typical magnitude of $\xi$ is 10--20\%, which would be a
reasonable size for an
${\cal O}(\alpha_s) = {\cal O}(\beta )$ correction.
This behaviour holds also true for 
$\xi$ at $p$ off the peak of the top-quark momentum distribution.
These features of the correction factor $\xi$ are consistent
with our qualitative argument in section 3.

We made a cross-check of our numerical results for $\xi$ by numerically
integrating 
$\xi \times \Gamma_t^{-1} d\Gamma_{t \to bl^+\nu}/ dx_l d\Omega_l$
over the
lepton energy-angular variables $\int dx_l d\Omega_l$ 
and comparing to the final-state interaction corrections to top-quark
momentum distribution given in Refs.\ \cite{r40,r26}.

It is seen that in Figs.~\ref{momdist} and 
\ref{angdist} the coordinate-space approach and the
momentum-space approach produce slightly different results.
In particular the normalization of the
cross sections differ at lower c.m.\ energies,
$0 \simlt \Delta E \simlt 2$~GeV, whereas the differences decrease at higher
energies.
The cause of the differences 
can be traced back to the different short-distance QCD
potentials employed in  the two approaches.
As we will discuss in section 7, this difference is formally counted
as higher order beyond our approximation, and
at present it should be taken as an uncertainty of the
theoretical prediction.

\section{Observable Proper to Top Decay Processes}
\cleqn
\clfn

As we have seen in the previous section, the final-state interactions
affect the $l^+$ distribution in top-quark decays.
The correction factor $\xi$ depends on the
kinematical variables of both the top quark and $l^+$, and
destroys the factorization of the cross section Eq.\ (\ref{born}).
In this section we define
an observable which depends
only on the top-quark decay process 
($d\Gamma_{t \to bl^+\nu}/ dx_l d\Omega_l$
of a free polarized top quark).\footnote{
This property is true only up to
${\cal O}(\alpha_s)={\cal O}(\beta )$ corrections
and may be violated by yet uncalculated 
${\cal O}(\alpha_s^2)$ corrections.
}
It is a differential quantity dependent
on the $l^+$ energy-angular variables.

  From Eq.\ (\ref{onepmint}) one sees that the correction 
factor $\xi$ is invariant
under the simultaneous transformations of the angular variables
\bea
\cos\theta_{Wl} \to -\cos\theta_{Wl} , ~~~~~
\cos\theta_{lt} \to -\cos\theta_{lt} ,
\label{trangles}
\eea
since $z_\pm$ is invariant.
This invariance may be understood as follows.
The soft-gluon factor,
which represents the final-state interaction
in diagram (b) (Fig.~\ref{fsifig}), 
does not depend on
the $l^+$ momentum as long as the $W^+$ momentum is kept fixed.
%because it is included in the hadronic sector and is independent of
%the spin of $b$ or $\bar{t}$.
One sees that accordingly the integrand of Eq.\ (\ref{xi1}) is
independent of the $l^+$ energy and angle.
The dependence on these variables enters only through the phase-space
integration over $\phi_{bl}$ for a given $l^+$ configuration.
Therefore, one may reverse the $l^+$ momentum in the $t$ rest frame
without affecting the phase-space integration, thereby keeping
the whole function $\xi$ also unchanged.
The above transformations Eq.\ (\ref{trangles}) are essentially this
reversal of the $l^+$ momentum.
(Due to the form of the integrand, there are extra degrees of 
freedom for the transformation of the $l^+$
direction, see below.)

Using Eq.\ (\ref{zangles}),
the above transformation can be written as a transformation
of the lepton energy and angle as
\bea
x_l \to x'_l = \left(
\frac{1+y}{y} - \frac{1}{x_l} 
\right)^{-1} , ~~~~~
\hat{\vc{n}}_l \to \hat{\vc{n}}_l' .
\eea
Here, $\hat{\vc{n}}_l = \vc{p}_l/|\vc{p}_l|$ denotes the unit vector
in the direction of $l^+$ in the top-quark rest
frame.
The choice of $\hat{\vc{n}}_l'$ for flipping the sign of $\cos\theta_{lt}$
$( \cos\theta_{lt} \to -\cos\theta_{lt}  )$
is not unique.
We represent by $\hat{\vc{n}}_l'$ an arbitrary one of those choices.
The production cross section of the top quark is not affected by 
this transformation since the top-quark kinematical
variables are not involved.
The important point is that neither the final-state interaction
correction is affected by it.

Now, using this invariance, we first construct a quantity that has the
simplest structure from the theoretical point of view, and afterwards we
present an improved quantity that will be more useful
for practical purposes.
Let us define
\bea
A( x_l, 
\hat{\vc{n}}_l \! \cdot \! \mbox{\boldmath$\cal P$}, 
\hat{\vc{n}}_l' \! \cdot \! \mbox{\boldmath$\cal P$} )
\equiv
\left[
\frac{d\sigma(\epem \! \to \ttbar \to bl^+\nu\bar{b}W^-)}
{d^3\vc{p}_t dx_l d\Omega_l}
\right]
\biggl/
\left[
\frac{d\sigma(\epem \! \to \ttbar \to bl^+\nu\bar{b}W^-)}
{d^3\vc{p}_t dx_l d\Omega_l}
\right]_{
\begin{array}{l}
{\scriptstyle x_l \to x'_l }\\
{\scriptstyle \hat{\vc{n}}_l \to \hat{\vc{n}}_l' }
\end{array}
} .
\nonumber \\
\label{defA}
\eea
The production cross section and the correction factor $\xi$
cancel in the numerator and denominator.
As a result, this quantity
is independent of the top-quark momentum $\vc{p}_t$
and is
determined only from the free polarized top-quark decay cross section.
In fact, using Eq.\ (\ref{fullformula}), we find that
\bea
&&
A( x_l, 
\hat{\vc{n}}_l \! \cdot \! \mbox{\boldmath$\cal P$}, 
\hat{\vc{n}}_l' \! \cdot \! \mbox{\boldmath$\cal P$} )
=
\left[
\frac{d\Gamma_{t \to bl^+\nu}
(\mbox{\boldmath$\cal P$}) }
{dx_l d\Omega_l} 
\right]
\biggl/
\left[
\frac{d\Gamma_{t \to bl^+\nu}
(\mbox{\boldmath$\cal P$}) }
{dx_l d\Omega_l} 
\right]_{
\begin{array}{l}
{\scriptstyle x_l \to x'_l }\\
{\scriptstyle \hat{\vc{n}}_l \to \hat{\vc{n}}_l' }
\end{array}
} 
\label{thA}
\eea
holds up to (and including) 
${\cal O}(\alpha_s)={\cal O}(\beta )$ corrections.
%Also, according to its construction, it has a transformation
%property
%\bea
%A( x_l, 
%\hat{\vc{n}}_l \! \cdot \! \mbox{\boldmath$\cal P$}, 
%\hat{\vc{n}}_l' \! \cdot \! \mbox{\boldmath$\cal P$} )
%=
%\frac{1}{A( x'_l, 
%\hat{\vc{n}}_l' \! \cdot \! \mbox{\boldmath$\cal P$}, 
%\hat{\vc{n}}_l \! \cdot \! \mbox{\boldmath$\cal P$} )
%} .
%\eea
Note that the polarization vector $\mbox{\boldmath$\cal P$}$, 
which specifies the decay
distribution, includes the correction induced
by the final-state interaction between $t$ and $\bar{b}$;
see Eq.\ (\ref{modpolvec}).

Since we take a ratio of differential cross sections in
the definition of 
$A$ in Eq.~(\ref{defA}),
this quantity would suffer from a large statistical error
experimentally.
Meanwhile, this quantity is predicted to be dependent only
on a few variables theoretically. (See Eq.~(\ref{thA}).)
It means that we may integrate out the irrelevant kinematical variables
before taking the ratio and reduce its statistical 
uncertainty.\footnote{
Consider a ratio of certain physical quantities which depend on a set of
kinematical variables
\bea
R = \frac{X( \phi_i)}{Y(\phi'_i)},
\nonumber
\eea
where $\phi_i$ denotes a point in the phase space, and $\phi'_i$ is obtained by
a transformation from it, $\phi'_i = \phi'(\phi_i)$.
Whenever the ratio $R$ takes the same value in 
a subspace $U$ of the whole phase space,
we may take sum over the subspace before taking the ratio:
\bea
R = \frac{\sum_{i \subset U} X( \phi_i)}
{\sum_{i \subset U} Y(\phi'_i)} .
\nonumber
\eea
}
For instance, we may choose $\hat{\vc{n}}_l' = - \hat{\vc{n}}_l$ and
define
\bea
\overline{A}( x_l, a)
\equiv
%\nonumber \\
\frac{\displaystyle
\int {d^3\vc{p}_t d\Omega_l}
\, \, \delta \!
\left( \hat{\vc{n}}_l \! \cdot \! \mbox{\boldmath$\cal P$} - a \right)
\left[
\frac{d\sigma(\epem \! \to \ttbar \to bl^+\nu\bar{b}W^-)}
{d^3\vc{p}_t dx_l d\Omega_l}
\right] ~~~~~
}{\displaystyle
%\biggl/
\int {d^3\vc{p}_t d\Omega_l}
\, \, \delta \!
\left( \hat{\vc{n}}_l \! \cdot \! \mbox{\boldmath$\cal P$} + a \right)
\left[
\frac{d\sigma(\epem \! \to \ttbar \to bl^+\nu\bar{b}W^-)}
{d^3\vc{p}_t dx_l d\Omega_l}
\right]_{x_l \to x'_l }
} .
%\label{}
\eea
Here, the top-quark polarization vector in the delta functions 
should be evaluated as
a function of $\vc{p}_t$ according to Eqs.~(\ref{thr_long})-(\ref{thr_norm}),
(\ref{diagrama3}) and (\ref{modpolvec}).
The numerator and denominator, respectively, depend 
on two external kinematical variables and all other 
variables are integrated out.

Again using Eq.\ (\ref{fullformula}), one finds that
$\overline{A}$ is determined solely from the free polarized top decays:
\bea
\overline{A}( x_l, a)
=
\left[
\frac{d\Gamma_{t \to bl^+\nu}
(\mbox{\boldmath$\cal P$}) }
{dx_l d\Omega_l} 
\right]_{\scriptstyle \hat{\vc{n}}_l \! \cdot \! 
\mbox{\boldmath$\scriptstyle{\cal P}$} = a} 
\biggl/
\left[
\frac{d\Gamma_{t \to bl^+\nu}
(\mbox{\boldmath$\cal P$}) }
{dx_l d\Omega_l} 
\right]_{
\begin{array}{l}
{\scriptstyle x_l \to x'_l }\\
{\scriptstyle \hat{\vc{n}}_l \! \cdot \! 
\mbox{\boldmath$\scriptstyle{\cal P}$} = -a}
\end{array}
} .
%\label{}
\eea
This is a general formula that is
valid even if the decay vertices of the top quark deviate
from the standard-model forms.\footnote{
Note that, quite generally,
energy-angular distributions of $l^+$ from free polarized top quarks
have a form
\bea
\frac{d\Gamma_{t \to bl^+\nu}
(\mbox{\boldmath$\cal P$}) }
{dx_l d\Omega_l} 
= F_0(x_l) + 
(\hat{\vc{n}}_l \! \cdot \! \mbox{\boldmath$\cal P$}) F_1(x_l) .
\nonumber
\eea
}
We see that the quantity $\overline{A}(x_l, a)$
preserves most of the differential information\footnote{
According to its construction, it satisfies a relation
\bea
\overline{A}( x_l, a )
\cdot
\overline{A}( x'_l, -a ) = 1 .
\nonumber
\eea
}
contained in
$d\Gamma_{t \to bl^+\nu}/ dx_l d\Omega_l$.

\section{Discussion}
\cleqn
\clfn

In this section we discuss three different issues relevant to our
work.
These are: 
the difference between the coordinate-space and momentum-space potentials,
the mis-assignment of the top-quark momentum, and
the disappearance of the final-state interaction corrections
at the various levels of inclusive cross sections.
\medbreak

As we have seen in section 5, our numerical results obtained from the
coordinate-space calculations and those obtained from the
momentum-space calculations differ slightly, although
all the qualitative features are common.
The difference can be traced back to the difference in the short-distance
part of the QCD potentials used in the two approaches.

Let us remind the reader how each potential is constructed
(in the short-distance regime).
The large-momentum part of the momentum-space potential $V_{\rm JKT}$
\cite{r35,r36} is determined as follows.
First the potential has been calculated up to the
next-to-leading order in a fixed-order calculation.
The potential is then improved using the
two-loop renormalization group equation in momentum space.
On the other hand, the short-distance part of the
coordinate-space potential $V_{\rm SFHMN}$ \cite{r9}
is calculated by taking the Fourier transform of the fixed-order potential
in momentum space, and
then the potential is improved using the two-loop renormalization group
equation in coordinate space.
Thus, the two potentials are {\it not}
the Fourier transforms of each other.
Only the leading and next-to-leading logarithmic terms of the series expansion
in a fixed $\overline{\rm MS}$-coupling are the same
for the two potentials.
The difference begins
at the next-to-next-to-leading order terms.
(The non-logarithmic term in the two-loop fixed-order correction.)

\begin{figure} \begin{center}
  \epsfxsize 80mm \mbox{\epsfbox{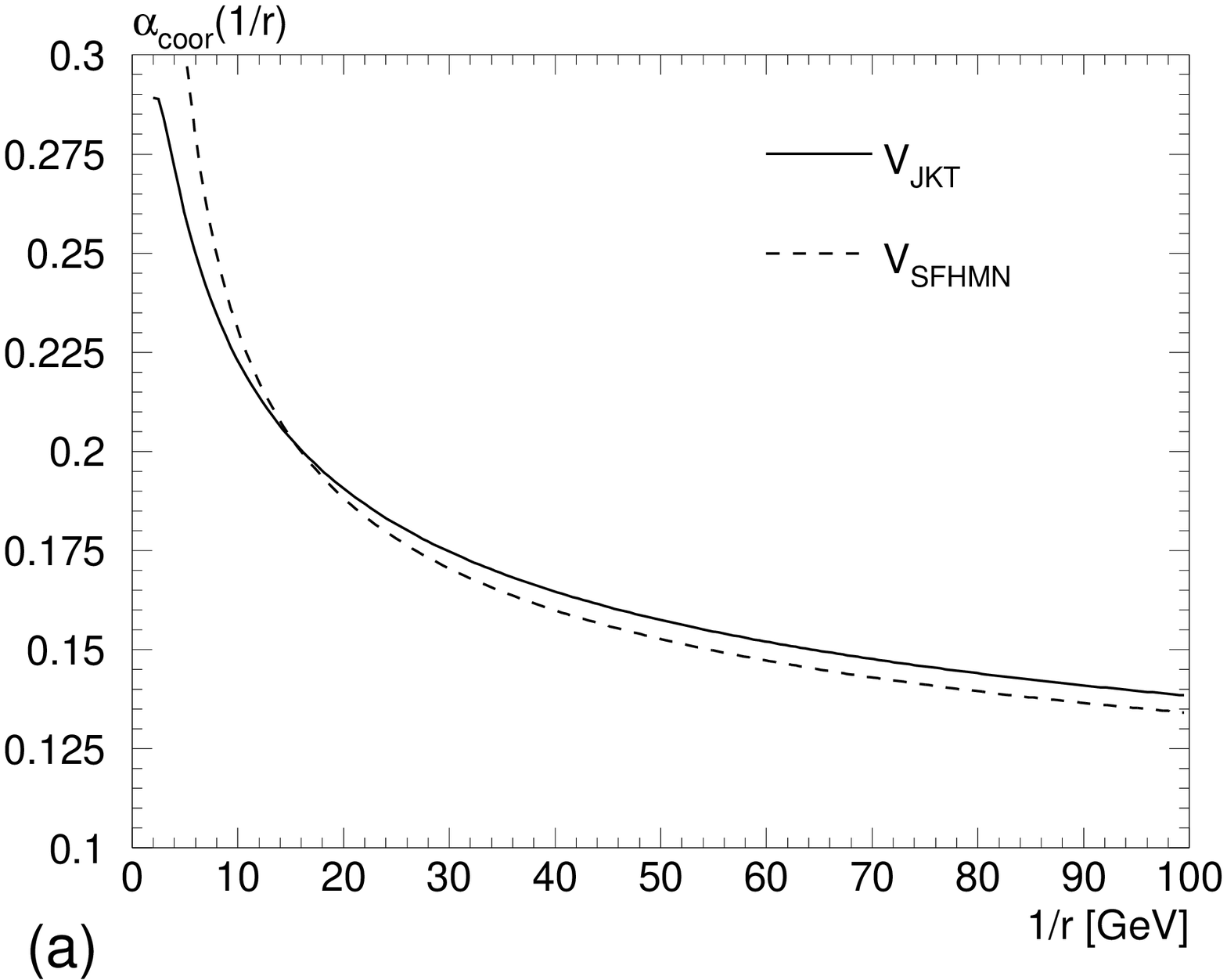}}
  \epsfxsize 80mm \mbox{\epsfbox{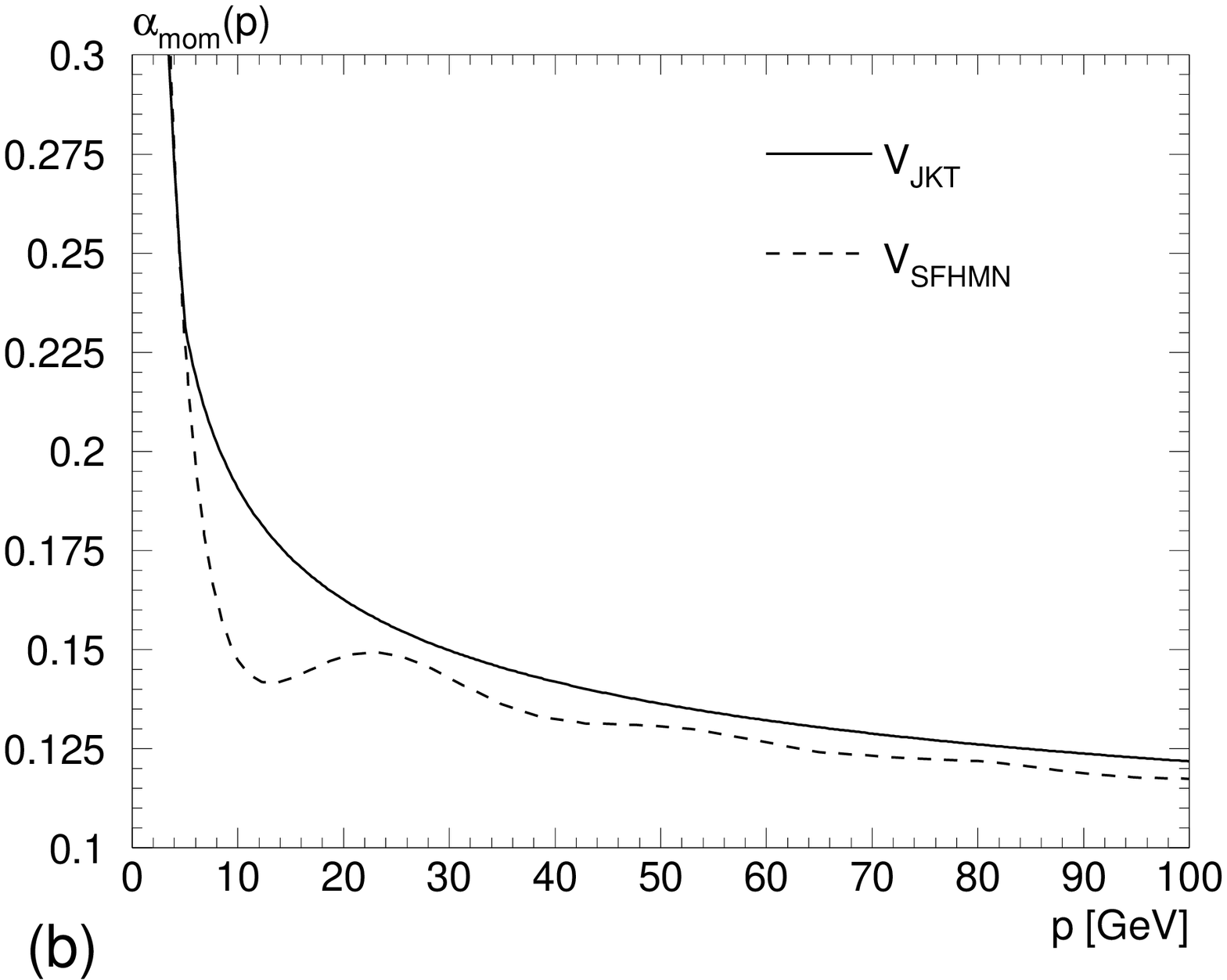}}
  \end{center}
  \caption[]{\label{runalf}The effective charges (a) $\alpha_{\mbox{\scriptsize
   coor}} ( 1/r )$ and (b) $\alpha_{\mbox{\scriptsize mom}} ( p )$ defined
  via the potentials in coordinate space and momentum space, respectively.
  $V_{\rm JKT}$ \cite{r35,r36} and $V_{\rm SFHMN}$ \cite{r9} represent the
  potentials used in the momentum-space approach and the coordinate-space
  approach, respectively.
  }
\end{figure}

To make a clear comparison, the two potentials are Fourier transformed 
numerically and 
we examine their difference
both in coordinate space and in momentum space.
We show the effective charges defined as 
\bea
\alpha_{\mbox{\scriptsize coor}} ( 1/r ) 
= \left( - C_F / r \right)^{-1} V(r) , ~~~~~
\alpha_{\mbox{\scriptsize mom}} ( q ) 
= \left( - 4 \pi C_F / q^2 \right)^{-1} \tilde{V}(q)
\eea
in Fig.~\ref{runalf}, which clearly demonstrates
that there is a non-negligible difference between the potentials.
The oscillatory behaviour of $V_{\rm SFHMN}$ 
in Fig.~\ref{runalf}(b) is an artifact due
to the discontinuity in the second derivative of the coordinate-space
potential,  which is located at the
continuation point of the perturbative potential to a long-distance
potential.
As already stated, the coordinate-space potential $V_{\rm SFHMN}$
follows the form required by
the two-loop renormalization group equation in the short-distance region, 
whereas the momentum-space potential $V_{\rm JKT}$ 
follows the form required by the 
two-loop renormalization group equation in the large momentum region.

In principle we may reduce the difference by including the two-loop
finite correction and invoking the three-loop renormalization group 
improvement \cite{mp}.
We shall not do so in this paper, because there are a number of other
corrections of the same order of magnitude which are not calculated yet.
We will study the difference of the two approaches in more detail
in a forthcoming paper.
\medbreak 

In subsection 4.c, we assumed a perfect assignment of the top-quark
momentum in cases (A) and (B) for defining the distribution formula,
Eq.\ (\ref{fullformula}).
In real experiments, however,
a mis-assignment of the top-quark momentum will be inevitable
whenever there is real gluon radiation in the final state
because of the typical jet-clustering algorithm that will be used.
For instance, when a gluon is indistinguishable from a $b$-jet, the 
top-quark momentum reconstructed by clustering
may be off-shell; rather grouping 
the gluon on the other side (with $\bar{b}$) would 
result in an on-shell momentum.\footnote{
There is no ambiguity in assigning the gluon to the production of \ttbar
since real gluon radiation in the top-quark production process 
is suppressed near threshold.}
Ref.\ \cite{r40} studied how this mis-assignment alters the top-quark
three-momentum distribution near threshold and found that the
correction is less than a few percent;
the clustering algorithm assumed in that paper, 
however, is somewhat unrealistic.
The effect of the mis-assignment was also studied in Ref.\ \cite{sch}
in the open-top region ($\sqrt{s} \gg 2m_t$) using a Monte Carlo
generator and with more realistic experimental assumptions.
It was shown that the effects on the top-quark invariant-mass
distribution and on the angular distributions are substantial at
$\sqrt{s} = 400$~GeV (for $m_t = 175$~GeV) and increase in 
magnitude and in complexity as the c.m.\ energy is raised.
Clearly, in our case, we need more 
detailed studies to see how a similar effect may influence our
results.
For this purpose, 
studies based on a Monte Carlo generator that produces the
fully-differential distribution including the full \oalfs corrections
near threshold would be necessary.
(See also Ref.\ \cite{khsj2} for analyses on the radiative color flows
from $t$, $\bar{t}$, $b$ and $\bar{b}$ close to the $t\bar{t}$ threshold.)
\medbreak

In the course of calculating the final-state interaction effects
on the lepton energy-angular distribution, we found that the final-state
interaction between $b$ and $\bar{b}$ vanishes (diagram (c) and (d))
if we add the corresponding
real-emission diagram as well as if we integrate over the top-quark energy.
This fact was already conjectured in Ref.\ \cite{r26} on account of an
estimate of the
Coulomb energy between $b$ and $\bar{b}$.
Nevertheless the same final-state interaction modifies the top-quark
energy distribution.
As a similar phenomenon one may be reminded of the cancellation of final-state
interactions (including also those between $t$ and $\bar{b}$) in the total
\ttbar production cross section, despite of the modification of the top-quark
three-momentum distribution.
In fact the cancellation of final-state interactions is a general
feature known in a wide class of inclusive hard scattering cross sections
both in QED and QCD.\footnote{
For example, it is found in Refs.\ \cite{my}-\cite{bcb} that the final-state
interaction correction to the invariant-mass distribution of $W$ changes
sign above and below the distribution peak and that the correction vanishes
upon integration over the invariant mass.
For an enlightening discussion on related problems, see 
also Ref.\ \cite{fkmc}.
}

\begin{table}
\begin{tabular}{c|ccc}   \hline
\rule[-3mm]{0mm}{8mm}
Inclusiveness           & Diagram (a)/(b)               & Diagram (c)   &
Diagram (d)+(e)*(f)
\\ \hline
\rule[-3mm]{0mm}{8mm}
fully differential      & $+$            & $+$            &
$+$              
\\
\rule[-3mm]{0mm}{8mm}
top energy integrated   & $+$            & vanish                &
vanish
\\
\rule[-3mm]{0mm}{8mm}
$\sigma_{\rm tot}(s)$& vanish                & vanish                &
vanish
\\ \hline \hline
\rule[-3mm]{0mm}{8mm}
typical gluon momentum  & $\alpha_s m_t$        & $\alpha_s^2 m_t$      &
$\alpha_s^2 m_t$
\\ \hline
\end{tabular}
\caption{
Vanishment or non-vanishment of final-state interaction diagrams
in the various inclusive cross sections.
The plus sign ``$+$'' shows there is a non-canceled
contribution from the diagram to the
corresponding cross section, while ``vanish'' shows the cancellation of
contributions from the diagram to the corresponding inclusive cross section.
For details, see subsection 4.b.
}
\end{table}

It may be worth summarizing here at which level of inclusiveness
the effects of the 
final-state interactions cancel in the various cross sections
in our particular process,
top quark pair production near threshold and its subsequent decay.
This is shown in Table~1.
Note that there are three typical mass scales involved in this process:
the top-quark mass $m_t$, the inverse Bohr radius $\alpha_s m_t$, and the
Coulomb energy between the \ttbar pair $\alpha_s^2 m_t$.
We show in the table the typical momentum scale of the gluon in
each diagram.

\section{Conclusion}
\cleqn

In this paper we 
considered the differential distribution of $l^+$ from the semi-leptonic
decay of the
top quark, where the parent top quark is
produced 
in $e^+e^- \to t\bar{t}$ near threshold.
Particularly, we have calculated the final-state interaction
corrections (rescattering corrections)
to the energy-angular distribution of leptons in the
top quark decay.
Also we have explicitly
written down the $l^+$ energy-angular distribution 
${d\sigma(\epem \! \to \ttbar \to bl^+\nu\bar{b}W^-)}/
{d^3\vc{p}_t dx_l d\Omega_l}$
including the full ${\cal O}(\alpha_s)={\cal O}(\beta )$ corrections
near threshold.

We presented numerical studies of the various effects of the final-state
interaction corrections.
All numerical results can be understood qualitatively from intuitive
pictures.
Attractive forces between $t$ and $\bar{b}$ and between $\bar{t}$ and $b$
modify not only the
momentum distribution of $bW^+$ or $\bar{b}W^-$ system but also
the top-quark polarization vector and 
the lepton energy-angular distribution.
\begin{itemize}
\item
The effect of Coulomb-gluon exchange between $\bar{b}$ and $t$,
when integrated over the $\bar{b}W^-$ phase-space,
can be regarded as a correction to the top-quark production process.
The effect can be incorporated by modifying the top-quark production
cross section and the top-quark polarization vector.
The top-quark momentum distribution is shifted to take a smaller
average momentum due to the attraction by $\bar{b}$.
Also since $\bar{b}$ is emitted preferably in the ${t}$ spin direction,
the attraction generates a $\cos \theta_{te}$ distribution of 
the top quark
as well as modifies the top-quark polarization vector.
\item
The Coulomb interaction between $b$ and $\bar{t}$ causes a 
non-factorizable correction with respect 
to the production and decay processes of the top quark.
It generates an energy-angle-correlated correction to the lepton
distribution.
Namely, the $l^+$ distribution is deformed in favour of the kinematical 
configurations
``small $E_l$ and emitted in $\bar{t}$ direction'' or
``large $E_l$ and emitted in $t$ direction'',
which can be understood as originating from the attraction of $b$ in the
direction of $\bar{t}$.
\item
Corrections from the gluon
exchange between $b$ and $\bar{b}$ turn out to vanish
when the top-quark energy is integrated out.
\end{itemize}

Without the non-factorizable effect $\xi$, the $l^+$ angular distribution
is dependent only on the polar angle from 
the polarization vector of the parent top quark
in its rest frame.
The final-state interaction brings in another direction into the problem,
the direction of $\bar{t}$, which is a completely new feature
in comparison to the decays of free polarized top quarks.

In order to study the decay properties of top quarks near $t\bar{t}$
threshold, it is desirable to extract the part which is specific to the
top-quark decay process alone.
In the case of semi-leptonic decay, we defined a quantity which
depends only on the decay distribution of a free polarized top quark.
The part which depends on $\cos\theta_{te}$ and $\cos\theta_{lt}$ is 
dropped using the transformation of the $l^+$ energy and angle
which leaves the final-state interaction unchanged.
Thus, we recover a differential quantity
$\overline{A}
( x_l, \hat{\vc{n}}_l \! \cdot \! \mbox{\boldmath$\cal P$})$
dependent only on the lepton energy and the lepton angle from the
parent top-quark polarization vector.

This quantity will be useful from the theoretical
point of view.
It can be calculated from the decay distribution of free top quarks
without including the bound-state effects or the final-state interaction 
corrections that are
typical to the threshold region.
Therefore, a variety of former studies on free top-quark decays may
also be applicable in the $t\bar{t}$ threshold region, where
highly polarized top quarks are available with the largest cross sections.
\medbreak

The authors wish to thank
M.~Je\.{z}abek and J.H.~K\"{u}hn for enlightening discussion.

\def\app#1#2#3{{\it Acta~Phys.~Polonica~}{\bf B #1} (#2) #3}
\def\apa#1#2#3{{\it Acta Physica Austriaca~}{\bf#1} (#2) #3}
\def\fortp#1#2#3{{\it Fortschr.~Phys.~}{\bf#1} (#2) #3}
\def\npb#1#2#3{{\it Nucl.~Phys.~}{\bf B #1} (#2) #3}
\def\plb#1#2#3{{\it Phys.~Lett.~}{\bf B #1} (#2) #3}
\def\prd#1#2#3{{\it Phys.~Rev.~}{\bf D #1} (#2) #3}
\def\pR#1#2#3{{\it Phys.~Rev.~}{\bf #1} (#2) #3}
\def\prl#1#2#3{{\it Phys.~Rev.~Lett.~}{\bf #1} (#2) #3}
\def\prc#1#2#3{{\it Phys.~Reports }{\bf #1} (#2) #3}
\def\cpc#1#2#3{{\it Comp.~Phys.~Commun.~}{\bf #1} (#2) #3}
\def\nim#1#2#3{{\it Nucl.~Inst.~Meth.~}{\bf #1} (#2) #3}
\def\pr#1#2#3{{\it Phys.~Reports }{\bf #1} (#2) #3}
\def\sovnp#1#2#3{{\it Sov.~J.~Nucl.~Phys.~}{\bf #1} (#2) #3}
\def\yadfiz#1#2#3{{\it Yad.~Fiz.~}{\bf #1} (#2) #3}
\def\jetp#1#2#3{{\it JETP~Lett.~}{\bf #1} (#2) #3}
\def\zpc#1#2#3{{\it Z.~Phys.~}{\bf C #1} (#2) #3}
\def\ptp#1#2#3{{\it Prog.~Theor.~Phys.~}{\bf #1} (#2) #3}
\def\nca#1#2#3{{\it Nouvo~Cim.~}{\bf #1A} (#2) #3}

%%%%% end of document 

\end{document}